\documentclass[showpacs,preprintnumbers,amsmath,amssymb,superscriptaddress]{revtex4}

\usepackage{graphicx}
\usepackage{subfigure}
\usepackage{dcolumn}
\usepackage{bm}
\usepackage{threeparttable}
\usepackage{multirow}
\usepackage{comment}
\usepackage{color}
\usepackage{pifont}
\usepackage[colorlinks,allcolors=blue]{hyperref}

\newcommand{\moe}{\affiliation{Key Laboratory of Atomic and Subatomic Structure and Quantum Control (MOE), Guangdong Basic Research Center of Excellence for Structure and Fundamental Interactions of Matter, Institute of Quantum Matter, South China Normal University, Guangzhou 510006, China
}}

\newcommand{\ihep}{\affiliation{Institute of High Energy Physics, Chinese Academy of Sciences, Beijing 100049, China}}

\newcommand{\iqm}{\affiliation{Guangdong-Hong Kong Joint Laboratory of Quantum Matter, Guangdong Provincial Key Laboratory of Nuclear Science, Southern Nuclear Science Computing Center, South China Normal University, Guangzhou 510006, China}}

\newcommand{\scnt}{\affiliation{Southern Center for Nuclear-Science Theory (SCNT), Institute of Modern Physics, Chinese Academy of Sciences, Huizhou 516000, Guangdong Province, China}}

\newcommand{\casu}{\affiliation{University of Chinese Academy of Sciences, Beijing 100049, China}}
\begin{document}
\title{\boldmath The pole structures of the $X(1840)/X(1835)$ and the $X(1880)$}

\author{Peng-Yu Niu}
\moe\iqm

\author{Zhen-Yu Zhang}
\moe\iqm

\author{Yi-Yao Li}
\moe\iqm


\author{Qian Wang}\email{qianwang@m.scnu.edu.cn}
\moe\iqm\scnt

\author{Qiang Zhao}\email{zhaoq@ihep.ac.cn}
\ihep\casu


\date{\today}

\begin{abstract}
Whether the $N\bar{N}$ interaction could 
form a state or not is a long standing question,
even before the observation of the $p\bar{p}$ threshold enhancement in 2003. The recent high statistic measurement in the $J/\psi \to \gamma 3(\pi^+\pi^-)$ channel would provide a good opportunity to probe the nature of the peak structures around the $p\bar{p}$ threshold in various processes. By constructing the $N\bar{N}$ 
interaction respecting chiral symmetry, we extract the pole positions by fitting the $p\bar{p}$ and $3(\pi^+\pi^-)$ invariant mass distributions of the $J/\psi \to \gamma p \bar p$ and $J/\psi \to \gamma 3(\pi^+\pi^-)$ processes. 
The threshold enhancement in the $p\bar{p}$ invariant mass distribution is from the pole on the third Riemann sheet, which more couples to the isospin triplet channel. The broader structure in the $3(\pi^+\pi^-)$ invariant mass comes from the pole on the physical Riemann sheet, which more couples to the isospin singlet channel. Furthermore, the large compositeness indicates that there 
should exit $p\bar{p}$ resonance based on the current experimental data. 
In addition, we also see a clear threshold enhancement in the $n\bar{n}$ channel, but not as significant as that in $p\bar{p}$ channel, which is useful and compared with further experimental measurement. 

\end{abstract}

\maketitle

\section{Introduction}

The study of bound states of proton-antiproton 
went back to 1949, before the establishment of quark model,
by Fermi and Yang~\cite{Fermi:1949voc}, 
who attempted to find the elementary particles in universe. 
In addition, Nambu and Jona-Lasinio also predicted~\cite{Nambu:1961tp,Nambu:1961fr}
the existence of a protonium with small binding energy. 
This kind of studies becomes more and more interesting until 
the observation of the proton-antiproton candidate,
i.e. the $X(1835)$ in 2003~\cite{BES:2003aic}, 
which is exactly the same year of the observation of the famous $X(3872)$ (or $\chi_{c1}(3872)$). Along this line,
experimentalists have put a great effort to search for 
proton-antiproton state in various processes,
for instance in the $\psi(nS) \to X A$ process, 
where $X$ is the observed $p\bar{p}$, $3(\pi^+\pi^-)$ channel 
and $A=\gamma,\pi,\eta,\rho,\omega,J/\psi$ is the spectator system~\cite{BES:2003aic,BES:2005ega,BES:2007inv,BESIII:2010vwa, CLEO:2010fre, BES:2009ufh, BESIII:2010gmv, BESIII:2011aa, BESIII:2013sbm, BESIII:2013lac,BESIII:2016fbr, BESIII:2023vvr}. The properties of the observed peak structures 
are collected in Tab.~\ref{tab:Xparticle}. 
 Meanwhile, the enhancement near the $p\bar p$ threshold is also observed in the other processes, e.g. $e^+ e^- \to p \bar p$~\cite{BaBar:2005pon, BaBar:2013ves, CMD-3:2015fvi}, the B decay sector~\cite{Belle:2002bro, Belle:2002fay, Belle:2007oni} and so on~\cite{BaBar:2007fsu, CMD-3:2013nph,Milstein:2022tfx}.
 However, the CLEO Collaboration does not find the $p\bar{p}$ 
 threshold enhancement in the $\psi(2S) \to X p \bar p$ process with 
 $X=\gamma, \pi^0, \eta$~\cite{CLEO:2010fre}. 
 This might because of the insufficient statistics.

A decade ago, the BESIII collaboration observed a peak at $1.84$ GeV, labeled as $X(1840)$, in the $3(\pi^+\pi^-)$ mass spectrum in the $J/\psi$ radiative decay process~\cite{BESIII:2013sbm}. Although the statistical significance of this structure is $7.6~\sigma$, further study should be performed to reveal whether this resonance peak is a new state or the  state in the $3(\pi^+\pi^-)$ decay mode. Recently, using more data, a further investigation on this $J/\psi$ radiative decay is performed~\cite{BESIII:2023vvr}. By fitting data, BESIII collaboration finds that the anomalous line shape in the $3(\pi^+\pi^-)$ spectrum around $1.84$ GeV could stem from the overlapping of the $X(1840)$ and the $X(1880)$ resonances (with the reduced $\chi^2$ as large as $\chi^2/N_{\mathrm{dof}}=155.6/41\approx3.8$)~\cite{BESIII:2023vvr}, 
 parameterized by Breit-Wigner formulae.
Their masses are $M(X(1840))=1832.5\pm 3.1\pm 2.5~\mathrm{MeV}$ and $M(X(1880))=1882.1\pm 1.7\pm 0.7~\mathrm{MeV}$, respectively. The widths are 
$\Gamma(X(1840))=80.7\pm 5.2\pm 7.7~\mathrm{MeV}$ and 
$\Gamma(X(1880))=30.7\pm 5.5\pm 2.4~\mathrm{MeV}$, respectively. 
However, the sum of their width ($111.4~\mathrm{MeV}$) is much larger than 
the distance between these two states ($49.6~\mathrm{MeV}$),
making the Breit-Wigner parameterization questionable. 
The former one is suggested to be consistent with 
the mass of the $X(1835)$ observed in the $\pi^+\pi^-\eta^\prime$ channel in $J/\psi$ radiative decay~\cite{BESIII:2016fbr}, but with much narrower width. Meanwhile, the fitting results of angular distributions of the $X(1840)$ and the $X(1880)$ show that both of them are pseudoscalar mesons~\cite{BESIII:2023vvr}. The latter one is considered as a new resonance with standard deviation larger than $10~\sigma$. However, as discussed above, the two broader states are obtained based on the Breit-Wigner formula, making the conclusion questionable. As the result, whether there should be one or two states in the $3(\pi^+\pi^-)$ channel is still an open question.

There are many interpretations~\cite{Gao:2003ka, Datta:2003iy, Yan:2004xs, Liu:2004er, Ding:2007xi, Deng:2012wi, Dedonder:2009bk, Liu:2009vm, Ding:2005gh, Wang:2006sna, Liu:2007tj,Wycech:2015qra} about the $p \bar p$ threshold enhancement, for instance 
final state interactions effects~\cite{Chen:2011yu, Chen:2010an, Chen:2008ee, Kang:2015yka, Milstein:2017dqp, Salnikov:2023ipo},
pseudoscalar glueballs~\cite{Li:2005vd, Kochelev:2005tu, Kochelev:2005vd, He:2005nm, Hao:2005hu, Gui:2019dtm},
radial excitations of the $\eta'$ meson~\cite{Huang:2005bc, Yu:2011ta,Wang:2020due} or a $3~^1S_0$ $q\bar q$ state~\cite{Li:2008mza} and other transition mechanism~\cite{Liu:2009vm}. 
For the details, we recommend the reviews Refs.~\cite{Liu:2016yfr,Ma:2024gsw}.
Although a great effort has been put forward,
the properties of the $X(1835)$ still remain controversial. 
The observations of the $X(1840)$ and the $X(1880)$ in the $3(\pi^+\pi^-)$ mass spectrum
of the $J/\psi \to \gamma 3(\pi^+\pi^-)$ process
may provide a golden opportunity to understand 
the $p\bar p$ threshold enhancement.

In this work, we are not intent on study all the abnormal enhancements, 
but only focus on the $J/\psi \to \gamma p \bar p$ and $J/\psi \to \gamma 3(\pi^+\pi^-)$ processes.
The paper is organized as follows:
our framework is presented in Sec.~\ref{sec:framework}. 
The results and discussions are presented in Sec.~\ref{sec:results}.
The summary and outlook are in the very end. 
Conventions used in this work and some useful formulas are presented in Appendix.

\begin{table}[h]
\renewcommand{\arraystretch}{1.3}
\caption{Summary of the properties of the $X(1835)$, the $X(1840)$ and the $X(1880)$. The mass and width of the $X(1835)$ are the average values provided by Particle Data Group(PDG)~\cite{Workman:2022ynf}. PDG use the data extracted from $J/\psi \to \gamma \pi^+\pi^- \eta'$~\cite{BESIII:2016fbr} and $J/\psi \to \gamma K^0_s K^0_s \eta$~\cite{BESIII:2015xco} for average. The properties of the $X(1840)$ and the $X(1880)$ are extracted from Ref.~\cite{BESIII:2023vvr}.}
\begin{tabular}{cccc}
\hline \hline
particle &$X(1835)$ &$X(1840)$ &$X(1880)$ \tabularnewline
\hline \hline
$J^{PC}$   &$0^{-+}$~\cite{BESIII:2011aa,BESIII:2015xco}&$0^{-+}$    &$0^{-+}$  \tabularnewline
mass (MeV)  &$1826.5^{+13.0}_{-3.4}$ &$1832.5\pm3.1\pm2.5$ & $1882.1\pm1.7 \pm 0.7$ \tabularnewline
width (MeV) &$242^{+14}_{-15}$      &$80.7\pm5.2\pm7.7$   & $30.7\pm5.5\pm2.4$ \tabularnewline
\hline \hline
\end{tabular}
\label{tab:Xparticle}
\end{table}

\section{Framework}
\label{sec:framework}

\begin{figure}[h]
\centering
\includegraphics[scale=0.5]{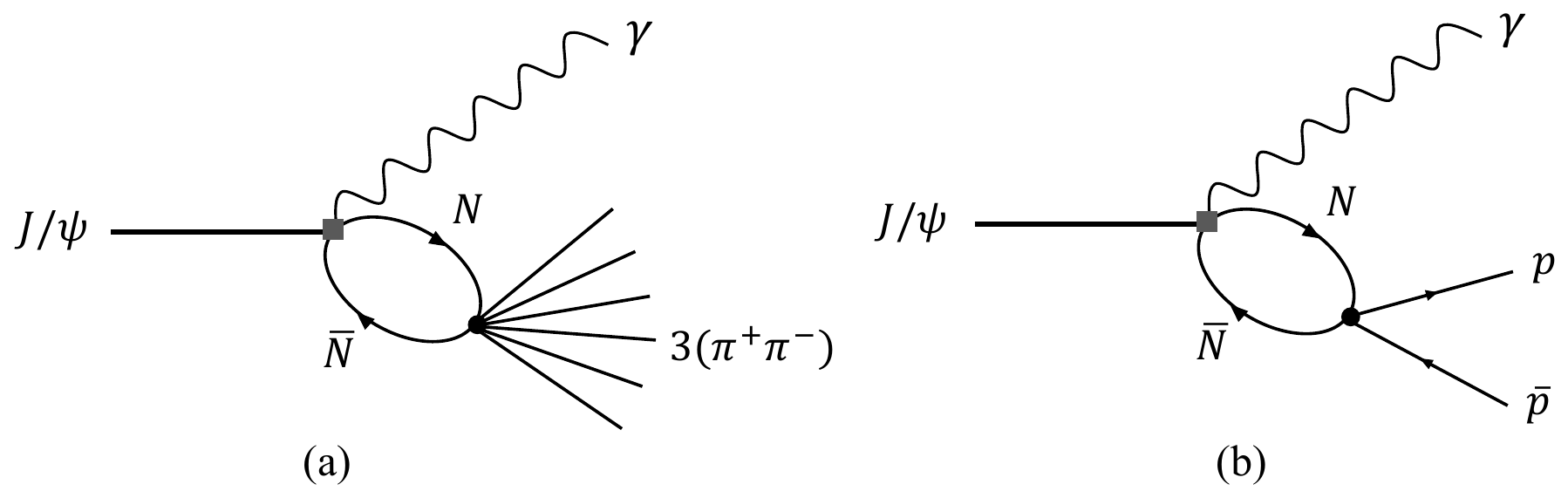} 
\caption{The schematic Feynman diagrams for the process $J/\psi\to \gamma 3(\pi^+\pi^-)$ (a) and $J/\psi \to \gamma p \bar p$ (b). }
\label{fig:fig2}
\end{figure}

As discussed in the above section, we aim at whether the dynamic generated states in 
the $N\bar{N}$ channel can describe the experimental 
data in both $J/\psi\to \gamma p\bar{p}$ and $J/\psi\to\gamma 3(\pi^+\pi^-)$ channels simultaneously or not. Accordingly, 
the $N\bar{N}$ and $3(\pi^+\pi^-)$ channels are considered dynamically and non-dynamically, respectively, in our framework. In this case, 
we would not include the contribution of  
resonances explicitly as that in Refs.~\cite{ Liu:2009vm,Sibirtsev:2004id}, but focus on the rescattering effect of the dynamic $N\bar N$ channel. The schematic diagrams are shown in Fig.~\ref{fig:fig2}. There are three kinds of vertexes, i.e. the dynamic $N \bar N\to N \bar N$ scattering amplitude, the $J/\psi \to \gamma N \bar N$ production amplitude, and the coupling between the dynamic $N\bar N$ channel and the non-dynamic $3(\pi^+\pi^-)$ channel.

\subsection{The $N\bar N$ effective potential and scattering amplitude}

The $N\bar N$ interaction has been widely discussed from different perspectives, including effective field theories~\cite{Chen:2011yu, Chen:2010an, Chen:2008ee}, chiral effective field theory~\cite{Kang:2013uia,Dai:2017ont,Xiao:2024jmu}, potential model~\cite{Cote:1982gr,Pignone:1994uj,El-Bennich:1999zbs, Datta:2003iy, Sibirtsev:2004id, Entem:2007bb, Ding:2007xi, Yan:2004xs, Dedonder:2009bk, Wycech:2015qra, Milstein:2017dqp, Salnikov:2023ipo} and so on. More discussions can be found in the review article~\cite{Richard:2022tpn}. The effective interaction of $N \bar N$ adopted here is constructed with respecting the chiral effective field theory. In the isospin limit, the effective interaction of $N \bar N$ contains two sectors, i.e. the contact potential and one(multi)-pion-exchanged potential. In our framework,
the leading and the next-leading order antinucleon-nucleon contact interactions are taken into account~\cite{Kang:2013uia,Dai:2017ont} as
%
\begin{align}
\label{eq:conV}
\mathcal{L}_{N N}^{(0)}&=C_S+C_T \bm \sigma_1\cdot \bm \sigma_2 ,\notag \\
\mathcal{L}_{N N}^{(2)}&=C_1 \bm q^2 +C_2 \bm k^2 +(C_3 \bm q^2 +C_4 \bm k^2)\bm \sigma_1\cdot \bm \sigma_2 +\frac{i}{2}C_5 (\bm \sigma_1+\bm \sigma_2)\cdot (\bm q\times \bm k) 
\notag \\
&+C_6 (\bm q\cdot\bm \sigma_1)(\bm q\cdot\bm \sigma_2)+C_7(\bm k\cdot\bm \sigma_1)(\bm k\cdot\bm \sigma_2),
\end{align}
where $C_i$ are low-energy constants (LECs) which need to be determined by fitting the experimental data. $\bm p$ and $\bm p'$ are the initial and final nucleon momenta in the center of mass system. $\bm k$ and $\bm q$ are defined by $\bm k=(\bm p' +\bm p)/2$ and $\bm q=\bm p'-\bm p$, respectively.

After performing the partial-wave projection, these terms contribute to both the S-wave $^1S_0$ and $^3S_1$ potentials as follow~\cite{Kang:2013uia,Dai:2017ont}:
\begin{align}
\label{eq:pwaV}
\mathcal{L}\left(^1 S_0\right) 
& =4 \pi\left(C_S-3 C_T\right)+\pi\left(4 C_1+C_2-12 C_3-3 C_4-4 C_6-C_7\right)\left(p^2+p^{\prime 2}\right) \notag \\
& =\tilde{C}_{^1S_0}+C_{^1S_0}\left(p^2+p^{\prime 2}\right), \\
\mathcal{L}\left(^3 S_1\right) 
& =4 \pi\left(C_S+C_T\right)+\frac{\pi}{3}\left(12 C_1+3 C_2+12 C_3+3 C_4+4 C_6+C_7\right)\left(p^2+p^{\prime 2}\right) \notag \\
& =\tilde{C}_{^3{S_1}}+C_{^3{S_1}}\left(p^2+p^{\prime 2}\right).
\end{align}
 While, in our convention, the terms contributed to the S-wave potential that are
\begin{align}
\mathcal{L}\left(^1 S_0\right) 
& =\left(C_S-3 C_T\right)+\frac{1}{4}\left(4 C_1+C_2-12 C_3-3 C_4-4 C_6-C_7\right)\left(p^2+p^{\prime 2}\right), \\ 
\mathcal{L}\left(^3 S_1\right) 
& =\left(C_S+C_T\right)+\frac{1}{12}\left(12 C_1+3 C_2+12 C_3+3 C_4+4 C_6+C_7\right)\left(p^2+p^{\prime 2}\right).
\end{align}
Compared with the partial wave decomposition formula in Ref.~\cite{Epelbaum:2004fk}, we have no factor of $4\pi$. The partial wave interaction can always be rewritten with
\begin{align}
\label{eq:swave}
\mathcal{L}\left(^1 S_0\right) 
& =C'_{01}+C'_{02}\left(p^2+p^{\prime 2}\right), \notag \\
\mathcal{L}\left(^3 S_1\right) 
& =C'_{11}+C'_{12}\left(p^2+p^{\prime 2}\right).
\end{align}
Here the LECs could be complex numbers~\cite{Chen:2010an,Dai:2017ont},
which stem from the annihilation mechanism (for instance a multi-pion system) for the $N\bar{N}$ system in comparison 
with the $NN$ system. 
%
%
On the other hand, the both isospin singlet and triplet are all allowed for each partial wave potentials for the $N\bar N$ system because of the absence of Pauli exclusion principle. As the spin and parity of the $X(1835)$ state is confirmed as a $0^{-+}$ state from experimental side~\cite{BESIII:2011aa, BESIII:2015xco}, 
we only consider the $^1S_0$ channel in the following. In addition, considering its isospin is not well determined in experiment
and isospin is not conserved in an electromagnetic process,
both $I=0$ and $I=1$ components may emerge 
in the $J/\psi$ radiative decays. 
As the result, the $N\bar N$ effective potentials in isospin basis adopted in this work are 
\begin{align}
V^{I}_{^1 S_0}=C_{I1}+C_{I2}\left(p^2+p^{\prime 2}\right),
\end{align}
where $I=0,1$. The relations between isospin basis and particle basis are 
\begin{align}
|I=1,I_3=0 \rangle=\frac{1}{\sqrt 2} \left(|p \bar p \rangle + |n \bar n \rangle\right),\quad 
|I=0,I_3=0 \rangle=\frac{1}{\sqrt 2} \left(|p \bar p \rangle - |n \bar n \rangle\right). 
\end{align}
With these relations, we can obtain the 
 $p \bar p \to p \bar p$, $n \bar n \to n \bar n$ and $p \bar p\to n \bar n $ potentials
\begin{align}
V_{p \bar p\to p \bar p}=V_{n \bar n\to n \bar n}= \frac{1}{2}\left( V^{I=1}+ V^{I=0} \right),\quad 
V_{p \bar p\leftrightarrow n\bar n }=\frac{1}{2}\left(V^{I=1}-V^{I=0} \right).
\end{align}
Here the $p\bar{p}$ and $n\bar{n}$ are denoted as the first and second channels, respectively.
One may notice that the $V_{p \bar p\to \bar n n }$ in Ref.~\cite{Dai:2017ont} is 
\begin{align}
V_{p \bar p\leftrightarrow n\bar n }=\frac{1}{2}\left(V^{I=0}-V^{I=1} \right),
\end{align} 
which stems from the different sign conventions of the isospin singlet and triplet wave functions. This only affect the sign of the off-diagonal elements in the potential and would not affect the physical results. As the result, the potential in particle basis read as
\begin{eqnarray}
V_{N\bar N\to N\bar N}(p,p')=
\frac{1}{2}\begin{pmatrix}
     C_{01}+C_{02}(p^2 +p^{\prime 2})+ C_{11}+C_{12}(p^2 +p^{\prime 2})   &  C_{01}+C_{02}(p^2 +p^{\prime 2})- C_{11}-C_{12}(p^2 +p^{\prime 2})\\
    C_{01}+C_{02}(p^2 +p^{\prime 2})- C_{11}-C_{12}(p^2 +p^{\prime 2})    & C_{01}+C_{02}(p^2 +p^{\prime 2})+ C_{11}+C_{12}(p^2 +p^{\prime 2})
    \end{pmatrix}.
\end{eqnarray}

The matrix of the two-body propagator is 
\begin{align} 
\mathcal{G}^+(E,\bm p)=\left[\begin{array}{cc}
G^+_{p\bar p}(E,\bm p)  & 0 \\
0  & G^+_{n\bar n}(E,\bm p)
\end{array}\right],
\end{align}
where $G_{N\bar N}(\bm p) $ is the two point function and expressed as
\begin{align}
G^+_{N\bar N}(E,\bm p)=\frac{1}{E-2m_{N}-\frac{\bm p^2}{2m_\mu}+i\epsilon^+}.
\end{align}
Here $E$ is the energy of the $N\bar N$ system and $m_\mu=m_{N}/2$ is the reduce mass of $N\bar N$ system.
The scattering amplitude 
$T_{N\bar N\to N\bar N}$ can be obtained by solving the Lippmann-Schwinger equation
\begin{align}
T_{N\bar N\to N\bar N}(p,p')=V_{N\bar N\to N\bar N}(p,p')+\int \frac{\mathrm{d}^3\bm p^{\prime\prime}}{(2\pi)^3}V_{N\bar N\to N\bar N}(p,\bm p'')\cdot \mathcal{G}^+ (E,\bm {p}^{\prime\prime}) \cdot T_{N\bar N\to N\bar N}(\bm p'',p').
\end{align}
The potential is the $N\bar N$ interaction potential that has been given above. We solve this integration equation numerically by discretizing momentum.

\subsection{The production amplitudes and differential widths}
As we try to understand whether the observed peak structures come from the dynamical rescattering of the $N\bar{N}$ interaction or not, we should not introduce any pole structures in the bare production amplitudes. 
In principle, the bare production amplitudes can be parameterized as polynomial of the invariant mass of the interested channel, for instance the quadratic polynomial form in Refs.~\cite{Yang:2022kpm, Dai:2018tlc}. Here we only keep the bare production amplitudes as contacts~\cite{Kang:2015hua}, i.e.
\begin{align}
\label{eq:transitionA}
\mathcal A^{0}_{J/\psi \to \gamma N \bar N}&=C_{J/\psi \to\gamma N \bar N}, \\
\mathcal A^{0}_{J/\psi \to \gamma 3(\pi^+\pi^-)}&=C_{J/\psi \to \gamma 3(\pi^+\pi^-)}
\end{align}
for the $J/\psi \to \gamma N\bar N$, $J/\psi \to \gamma 3(\pi^+\pi^-)$ processes. The $N \bar N \to 3(\pi^+\pi^-)$ transition amplitude is also parameterized as a constant
\begin{align}
\label{eq:transitionB}
\mathcal A_{N\bar N\to 3(\pi^+\pi^-)}&=C_{N\bar N\to 6\pi}.
\end{align}
In this case, the energy dependencies on the $p \bar p$ and $3(\pi^+\pi^-)$ invariant mass spectrum come 
solely from the $N\bar{N}$ Final State Interaction (FSI). 
With the pieces mentioned above, the Lorentz invariant physical decay amplitudes are expressed as~\cite{Dai:2018tlc,Yang:2022kpm}
\begin{align}
\tilde {\mathcal M}_{J/\psi \to \gamma p \bar p}&=
8\pi^2\sqrt{E_{J/\psi}E_{\gamma} E_{p}E_{\bar p}}\mathcal M_{J/\psi \to \gamma p \bar p},
\\  
\tilde {\mathcal M}_{J/\psi \to \gamma 3(\pi^+\pi^-)}  
&=32 \pi^{7/2} \sqrt{E_{J/\psi}  E_\gamma  E_{2}E_{3}E_{4} } \mathcal M_{J/\psi \to \gamma 3(\pi^+\pi^-)},
\end{align}
where
\begin{align}
\mathcal M_{J/\psi \to \gamma N \bar N}&= \mathcal A^{0}_{J/\psi \to\gamma N \bar N}+ \int \frac{\mathrm{d}^3\bm p}{(2\pi)^3} \mathcal A^{0}_{J/\psi \to\gamma N \bar N}\cdot\mathcal{G}^+\cdot T_{N\bar N\to N\bar N}, \\
\mathcal M_{J/\psi \to \gamma 3(\pi^+\pi^-)}&=\mathcal A^{0}_{J/\psi \to \gamma 3(\pi^+\pi^-)}+ \int \frac{\mathrm{d}^3\bm p}{(2\pi)^3}\mathcal M_{J/\psi \to \gamma N \bar N} \cdot \mathcal{G^+} \cdot \mathcal A_{N\bar N\to 3(\pi^+\pi^-)}.
\end{align}
$T_{N\bar N\to N\bar N}$ is the two-dimension transition matrix of $N\bar N\to N\bar N$ in the particle basis obtained in the last subsection. Here the momentum dependence of the $\mathcal M$, $\mathcal G^+$ and $T$ are implicit. 
The $E_{X}$ is the energy of the particle $X$ in the rest frame of $N\bar N$ and the definition of $E_{i}(i=2,3,4)$ are given later. 

The experimental data are extracted from Refs.~\cite{BES:2003aic,BESIII:2011aa} for $J/\psi \to \gamma p  \bar p$ and Refs.~\cite{BESIII:2023vvr} for ${J/\psi \to \gamma 3(\pi^+\pi^-)}$. For the latter one, the contribution of $J/\psi \to \pi^0 3(\pi^+ \pi^-)$ to the background is removed and the efficiency correction is considered. The dominant background for $J/\psi \to \gamma p \bar p$ is from $J/\psi \to \pi^0 p \bar p$~\cite{BES:2003aic,BESIII:2011aa} and can be described well by 
\begin{align}
f_{\mathrm{bkg}}(\delta)=a_1\delta^{1/2}+a_2\delta^{3/2}+a_3\delta^{5/2},
\end{align}
where $\delta=M_{p\bar p}-2m_p$~\cite{BES:2003aic}. Besides that, the $J/\psi \to \pi^0 p \bar p$ channel  was also studied by BESIII collaboration, which indicates that there is no evidence of a $p\bar p$ mass-threshold enhancement~\cite{BES:2009ufh,BESIII:2017qwj}.
While the background for the $J/\psi \to \gamma 3(\pi^+\pi^-)$ stems from more than one source. With the results given in Ref.~\cite{BESIII:2023vvr}, we assume that the background can be described with  a second-order polynomial function
\begin{align}
bg_{J/\psi \to \gamma 3(\pi^+\pi^-)}(Q)=a+ b Q+c Q^2
\end{align}
with $Q$ the invariant mass of $N\bar N$ or $3(\pi^+\pi^-)$ system. In our fitting processes, only the contribution of background for the process $J/\psi \to \gamma 3(\pi^+\pi^-)$ is included and is treated as an incoherent contribution.

For the $J/\psi$ three body decay process $J/\psi \to \gamma N \bar N$, the differential width can be obtained with 
\begin{align}
\frac{d \Gamma_{J/\psi \to X N  \bar N}}{dm_{N \bar N}}=\frac{\lambda^{1/2}(M_{J/\psi}^2,m_{N \bar N}^2,m_X^2)\sqrt{m_{N \bar N}^2-4m_N^2}}{2^7\pi^3 M_{J/\psi}^3}\frac{1}{2S_{J/\psi} +1} |\tilde {\mathcal M}_{J/\psi \to X N \bar N}|^2,
\end{align}
where $m^2_{N \bar N}=(p_N+p_{\bar N})^2$. Considering the complexity of the seven body phase space, $\pi^+ \pi^-$ can be treated as a whole particle with the mass $2m_\pi$~\cite{Yang:2022kpm} for the process $J/\psi \to \gamma 3(\pi^+ \pi^-)$. The masses of photon and the 3 quasi-particles are labeled with $m_1=m_\gamma=0$ and $m_2=m_3=m_4=2m_\pi$. Accordingly the energies of the 3 quasi-particles are labeled with $E_i(i=2,3,4)$. The differential decay width for the ``four-body" decay processes $J/\psi \to \gamma 3(\pi^+\pi^-)$ reads as 
\begin{align}
d\Gamma_{J/\psi \to \gamma 3(\pi^+\pi^-)}&=\frac{1}{2M_{J/\psi}} \frac{1}{\tilde N}\frac{1}{2S_{J/\psi}+1} \left[|\tilde {\mathcal M}_{J/\psi \to \gamma 3(\pi^+\pi^-)}|^2+ bg_{J/\psi \to \gamma 3(\pi^+\pi^-)}^2(m_{234}) \right] d\Phi_4.
\end{align}
where $\tilde N=3!$ is the factor caused by the identical particles. $m_{234}$ is the invariant mass of the six pions system.
With the four body phase space formula given in App.~\ref{app:kinematics}, the differential decay widths can be obtained with~\cite{Dai:2018tlc,Yang:2022kpm}
\begin{align}
\frac{d\Gamma^{\mathrm{total}}_{J/\psi \to \gamma 3(\pi^+\pi^-)}}{dm_{234}}&=\frac{d\Gamma_{J/\psi \to \gamma 3(\pi^+\pi^-)}}{dm_{234}}+\frac{d\Gamma^{\mathrm{bg}}_{J/\psi \to \gamma 3(\pi^+\pi^-)}}{dm_{234}},
\end{align}
where
\begin{align}
\frac{d\Gamma_{J/\psi \to \gamma 3(\pi^+\pi^-)}}{dm_{234}}&=\frac{1}{2^6 (2\pi)^5}\frac{M^2_{J/\psi}-m^2_{234}}{ m_{234} M^3_{J/\psi}}\frac{1}{2S_{J/\psi }+1}\frac{1}{\tilde N} \int_{s_{23}^-}^{s_{23}^+} d s_{23} \int_{s_{34}^-}^{s_{34}^+} d s_{34}|\tilde{\mathcal M}_{J/\psi \to \gamma 3(\pi^+\pi^-)}|^2, \\
\frac{d\Gamma^{bg}_{J/\psi \to \gamma 3(\pi^+\pi^-)}}{dm_{234}}&=\frac{1}{2^6 (2\pi)^5}\frac{M^2_{J/\psi}-m^2_{234}}{ m_{234} M^3_{J/\psi}}\frac{1}{2S_{J/\psi }+1}\frac{1}{\tilde N} \int_{s_{23}^-}^{s_{23}^+} d s_{23} \int_{s_{34}^-}^{s_{34}^+} d s_{34} bg^2(m_{234}).
\end{align}
The details of the upper and lower limits of the integral can be found in App.~\ref{app:kinematics}.

In order to fit the events distributions in experiment the fitting functions are written as follows. For the $J/\psi \to \gamma p  \bar p$ process, the fitting functions are 
\begin{align}
\mathrm{Events}(m_{p\bar p})&=\mathrm{fac1}\times \frac{d \Gamma_{J/\psi \to \gamma p\bar p}}{dm_{p \bar p}}, \\
\mathrm{Events}(m_{p\bar p})&=\mathrm{fac2}\times \frac{d \Gamma_{J/\psi \to \gamma p\bar p}}{dm_{p \bar p}}.
\end{align}
for the data sets from BES collaboration in 2003~\cite{BES:2003aic} and 2012~\cite{ BESIII:2011aa}, respectively. 
For the $J/\psi \to \gamma 3(\pi^+\pi^-)$ process, the fitting function is 
\begin{align}
\mathrm{Events}(m_{6\pi})&=\mathrm{fac3}\times \frac{d\Gamma_{J/\psi \to \gamma 3(\pi^+\pi^-)}}{dm_{6\pi}} + \frac{d\Gamma^{\mathrm{bg}}_{J/\psi \to \gamma 3(\pi^+\pi^-)}}{dm_{6\pi}}.
\end{align}

\section{Results and discussion}
\label{sec:results}
\begin{figure}[h]
\centering
\includegraphics[scale=0.5]{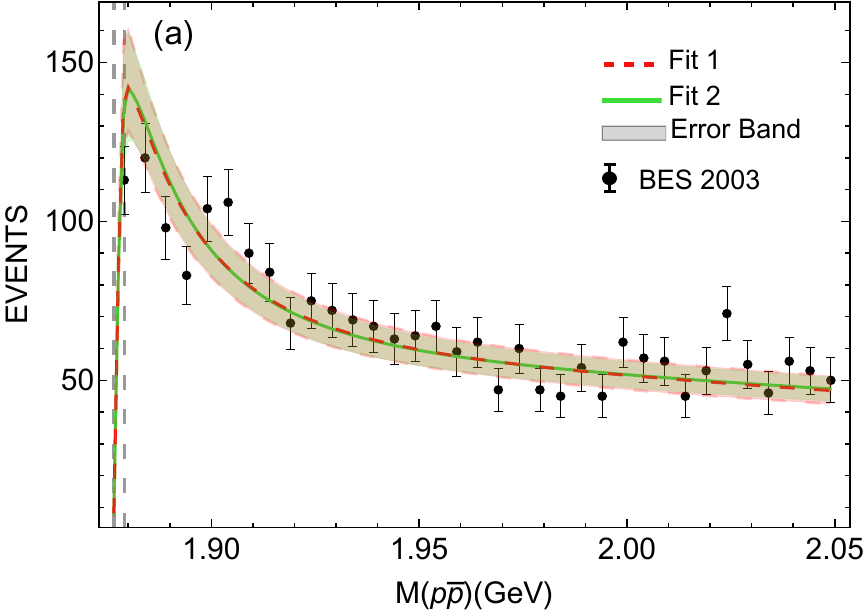} 
\includegraphics[scale=0.5]{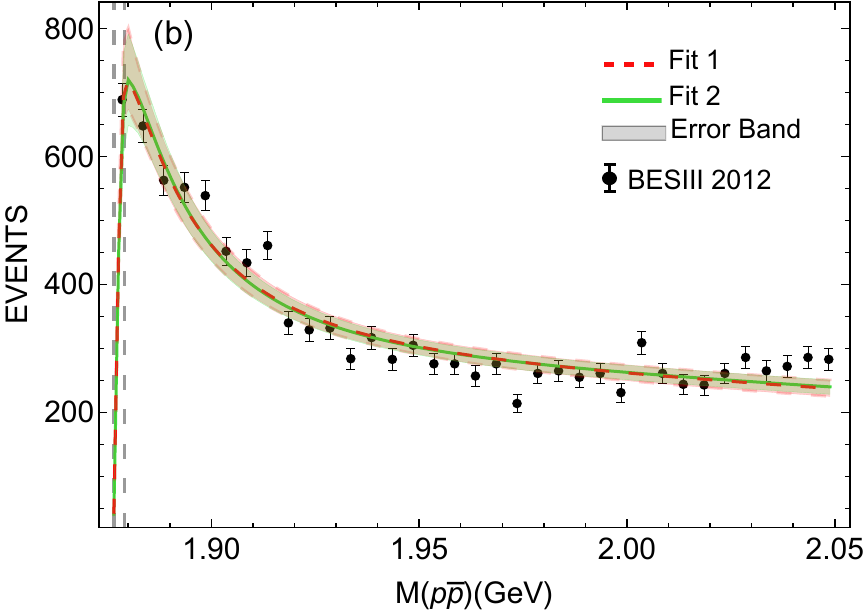} \\
\includegraphics[scale=0.5]{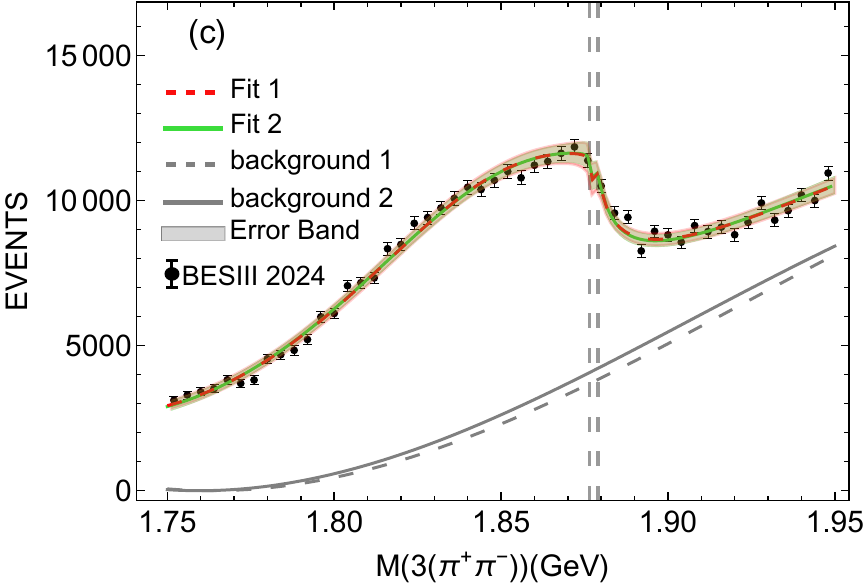}
\caption{The fitted line shapes for Solution-I (red dashed curve) and Solution-II (green solid curve). These two curves almost overlap.
The red and green bands are the one standard deviation uncertainties. The two vertical gray dashed lines are the $p\bar{p}$ and $n\bar{n}$ thresholds from left to right. 
The dashed and solid gray curves in (c) are the background contributions. The experimental data are taken from Refs.~\cite{BES:2003aic,BESIII:2011aa,BESIII:2023vvr}.}
\label{fig:lineshape}
\end{figure}

We use all the available experimental data, i.e. the data from both BESII~\cite{BES:2005ega} and BESIII~\cite{BESIII:2011aa, BESIII:2023vvr}, to extract our parameters. We perform about 300 times fits with Minuit2~\cite{James:1975dr} and find the best two solutions as presented in Tab.~\ref{tab:fitresults}. To take into account the annihilation contribution, the $C_{01}$ and $C_{11}$ parameters allow non-zero imaginary part. The corresponding lineshapes with error band are illustrated in Fig.~\ref{fig:lineshape}.
Although the two solutions can describe the experimental data almost equally well, the obtained parameters. For instance, the parameter fac3 of the solution-II is larger one order than that of the solution-I. This is because that the $C_{N\bar N \to 6 \pi}$ of the solution-I is about triple larger that of the solution-II, making the two solutions have the same event distribution for the process $J/\psi \to \gamma 3(\pi^+ \pi^-)$.

\begin{table}[h]
\renewcommand{\arraystretch}{1.3}
\caption{The fitted parameters of the two best solutions are presented in this table. The  parameter
 $\mathcal A^0_{J/\psi \to \gamma 3(\pi^+\pi^-)}$ process is set to be $1~\mathrm{GeV}^{-7/2}$ . The cut off in the integral equation is set to be $1~\mathrm{GeV}$. The reduced $\chi^2$ are listed in the last line and the value given in parenthesis is the reduced $\chi^2$ for the $J/\psi \to \gamma 3(\pi^+ \pi^-)$ channel. }
\begin{tabular}{rcc}
\hline \hline
Parameters & Solution-I & Solution-II \tabularnewline
\hline 
$C_{01}~(\mathrm{GeV}^{-2})$        &$(87.41\pm 0.32)-(6.08\pm 0.10)i$ &$(97.24\pm 0.66)-(-6.72\pm 0.16)i$ \tabularnewline
$C_{02}~(\mathrm{GeV}^{-4})$        &$-102.36\pm0.21$ &$-109.26\pm 0.47$  \tabularnewline
$C_{11}~(\mathrm{GeV}^{-2})$        &$(-33.47\pm0.31)+(0.56\pm 0.76)i$  &$(153.79\pm3.98 )+(12.56\pm 8.26)i$ \tabularnewline
$C_{12}~(\mathrm{GeV}^{-4})$        &$57.56\pm 16.44$   &$247.88\pm 2.01$   \tabularnewline
$C_{J/\psi \to \gamma p\bar p}~(\mathrm{GeV}^{-2})$     &$168.06\pm9.08$    &$-88.22\pm23.26$  \tabularnewline
$C_{J/\psi \to \gamma n\bar n}~(\mathrm{GeV}^{-2})$     &$-372.26\pm 8.23$  &$428.47\pm 8.39$\tabularnewline
$C_{p\bar p\to 6\pi}~(\mathrm{GeV}^{-7/2})$ &$-330.66\pm 10.63$  &$-111.08\pm 2.12$ \tabularnewline
$C_{n\bar n\to 6\pi}~(\mathrm{GeV}^{-7/2})$ &$263.04\pm 10.75$   &$81.65\pm 5.83$\tabularnewline
fac1 ($10^{-3}$)          &$1.42\pm 0.10$   &$0.85\pm 0.048$  \tabularnewline
fac2 ($10^{-3}$)          &$7.19\pm 0.50$    &$4.32\pm 0.23$ \tabularnewline
fac3 ($10^{-3}$)          &$0.71\pm 0.04$    &$3.14\pm 0.11$\tabularnewline
$a~(\mathrm{GeV}^{-1})$    &$-2.58\times 10^7\pm 1.61\times 10^4$ &$-2.84\times 10^7\pm 9.47\times 10^5$ \tabularnewline
$b~(\mathrm{GeV}^{-2})$    &$2.54\times 10^7\pm 8.50\times 10^3$ &$2.84\times 10^7\pm 1.12\times 10^3$ \tabularnewline
$c~(\mathrm{GeV}^{-3})$    &$-6.17\times 10^7\pm 4.38\times 10^3$   &$-6.96\times 10^6\pm 2.62\times 10^4$\tabularnewline
$\chi^2/\mathrm{d.o.f}$ & $2.32(2.24)$ & 2.33(2.31) \tabularnewline
\hline \hline
\end{tabular}
\label{tab:fitresults}
\end{table}

\begin{figure}[h]
\centering
\includegraphics[scale=0.8]{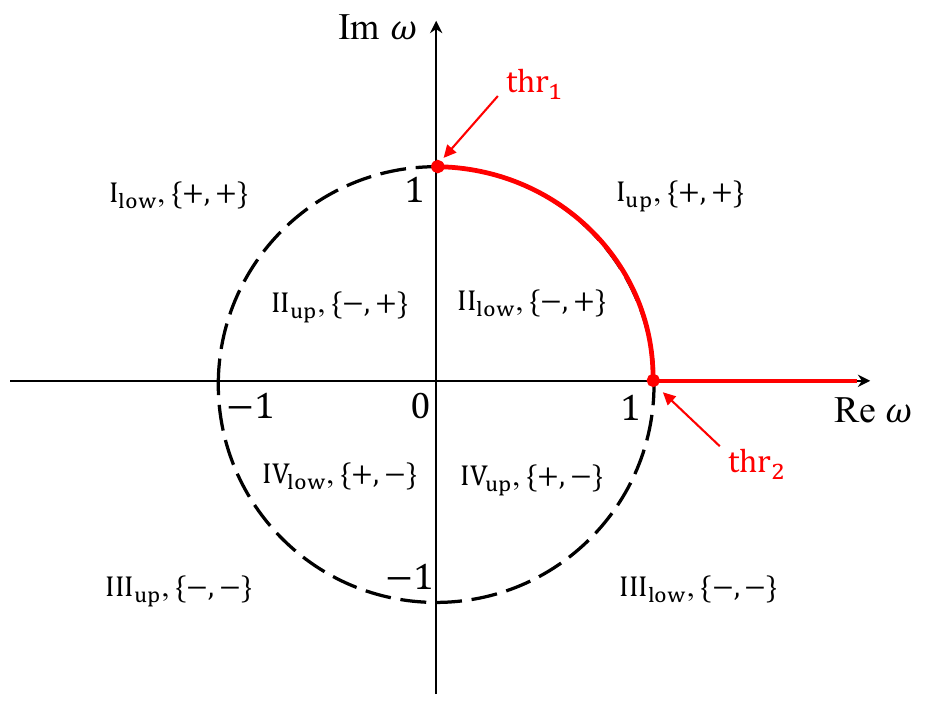} 
\caption{The $\omega$ plane for the two elastic channels. The $\omega$ plane is divided eight parts by the unit circle and coordinate axis. The four Riemann sheets are also labeled with I-IV. The red solid line indicates the unit cut on the energy plane. $\{+,+\}$ is the first Riemann sheet i.e. the physical Riemann sheet. $\mathrm{thr}_i$ is the threshold of the $i$th channel. The lower index ``up" and ``low" mean the upper and lower half plane of the corresponding Riemann sheet.}
\label{fig:oplane}
\end{figure}

With the fitted parameters, we can further extract the interested physical quantities, for instance pole positions, effective couplings and so on. 
For the two-channel case, there are four Riemann sheets which 
is labeled as $\{l,j\}$ with $l,j=+~\mathrm{ or }~-$ for 
the sign of the imaginary part of momentum of the $i$-th channel. According to the Riemann surface theory, these four Riemann sheets can be mapped onto the $\omega$ plane~\cite{Kato:1965iee,Badalian:1981xj,Baru:2019xnh, ParticleDataGroup:2022pth}, as illustrated by Fig.~\ref{fig:oplane}. The mapping relations are
\begin{align}
\label{eq:kk}
k_1=\sqrt{\frac{\mu_1 \Delta}{2} } \left( \omega+\frac{1}{\omega} \right),~
k_2=\sqrt{\frac{\mu_2 \Delta}{2} } \left( \omega-\frac{1}{\omega} \right),
\end{align}
where $\mu_i$ is the reduce mass of the $i$th channel and $\Delta=\mathrm{thr}_2-\mathrm{thr}_1$ is the distance between the two channels. The two channels are ordered by their thresholds, i.e. $p\bar{p}$ and $n\bar{n}$ channels are denoted as the first and the second channels, respectively.
With this mapping, one can search for pole on $\omega$-plane instead of energy plane. The energy can be obtained with the relation
\begin{eqnarray}
    E=\frac{k_1^2}{2\mu_1}=\frac{k_2^2}{2\mu_2}+\Delta=\frac{\Delta}{4}\left(\omega^2+\omega^{-2}+2\right).
\end{eqnarray}

Figure.~\ref{fig:oplane} shows that how the different Riemann sheet connects with each other on the $\omega$ plane. Switch to the energy plane, the lower half plane of the $\{-,-\}$ sheet is connected to the upper half plane of the $\{+,+\}$ sheet along the real axis $[\mathrm{thr}_2,+\infty)$. The lower half plane of the $\{-,+\}$ sheet connects to the upper half plane of the $\{+,+\}$ sheet along the real axis between $[\mathrm{thr}_1, \mathrm{thr}_2)$. $\mathrm{thr}_1$ and $\mathrm{thr}_2$ are the thresholds of the first and second channels, respectively. For the $N\bar N$ scattering, these two thresholds are very close to each other with distance $\Delta=2m_n-2m_p\approx 2.59$ MeV. Thus the poles on the Riemann sheet $\{-,+\}$ have limit effects on the physical observable unless they locate within the region $[\mathrm{thr}_1, \mathrm{thr}_2]$ which is a very narrow range compared with the whole measurement range.

\begin{figure}[h]
\centering
\includegraphics[scale=0.52]{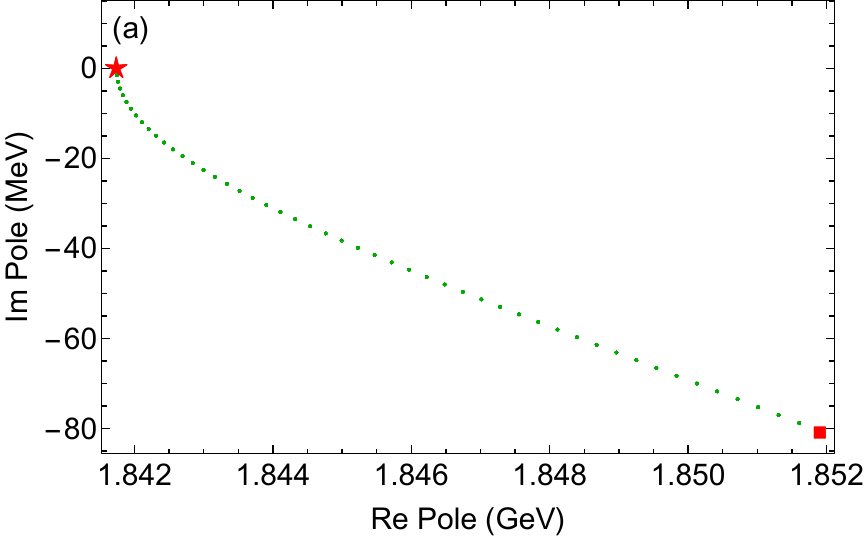}
\includegraphics[scale=0.5]{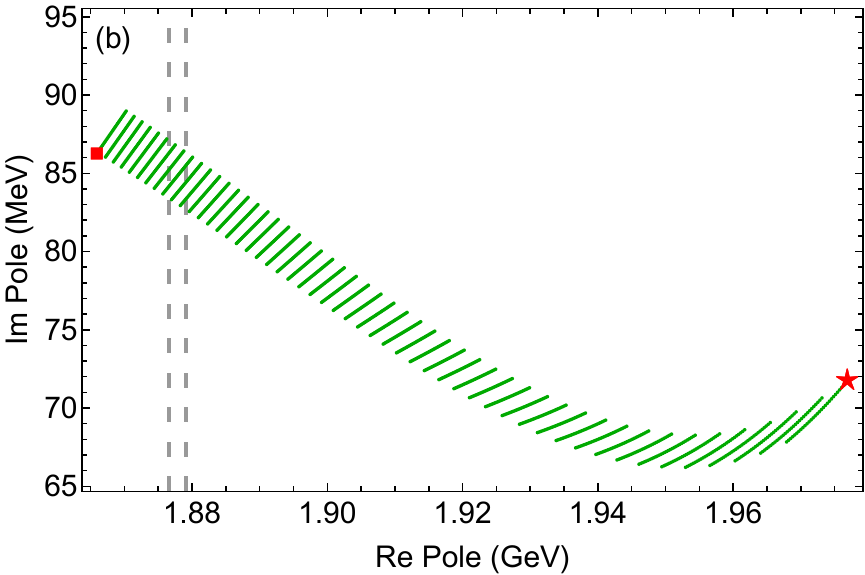}\\
\includegraphics[scale=0.5]{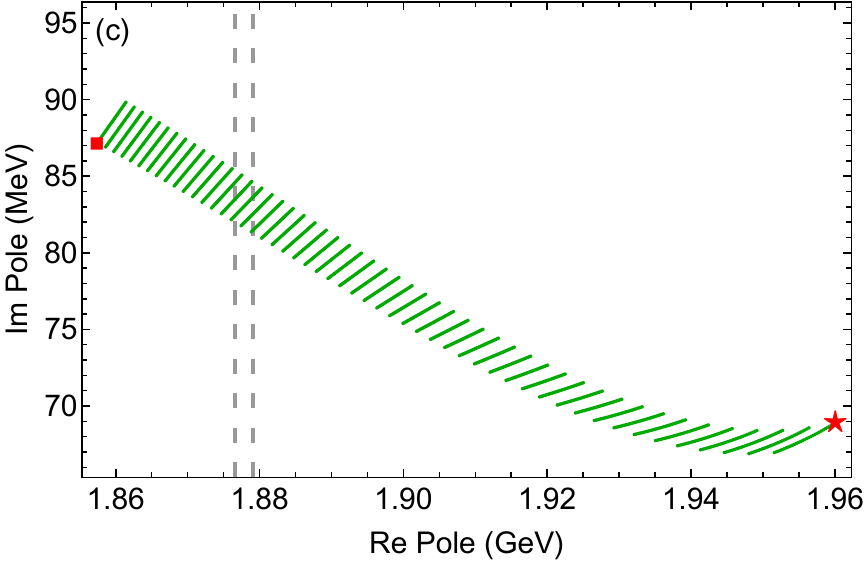}
\includegraphics[scale=0.5]{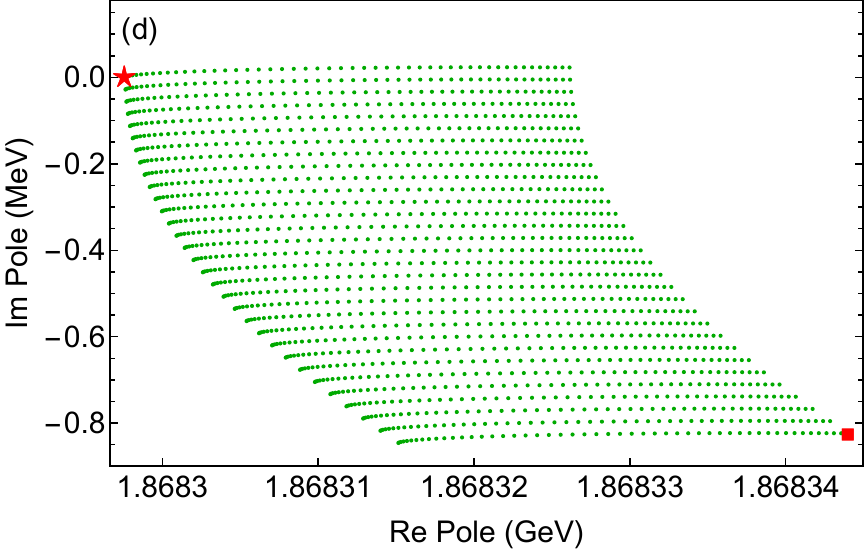}
\caption{The poles trajectories on the complex energy plane with the variation of $\mathrm{Im}~C_{01}$ and $\mathrm{Im}~C_{11}$ from the fitted values to zero in Solution-I. The red square and star label the initial and final position (i.e.  $\mathrm{Im}~C_{01}=\mathrm{Im}~C_{01}=0$), respectively. For Solution-II, the behaviors are similar to those of Solution-I. The two vertical dashed lines indicate the $p\bar{p}$ and $n\bar{n}$ thresholds.}
\label{fig:polePosition}
\end{figure}

The obtained poles are collected in Tab.~\ref{tab:pole}.
One can see that all of them are below the lowest threshold,
as the LECs $C_{01}$ and $C_{11}$ are set to be complex numbers. 
The reasonability of these poles can be seen by the pole trajectory, i.e. Fig.~\ref{fig:polePosition}, when the imaginary parts of the two LECs $C_{01}$ and $C_{11}$ go to zero. When $\mathrm{Im }~C_{01} \to 0$ and $\mathrm{Im }~C_{11} \to 0$, the poles on the first and third Riemann sheet will move to the real axis and below the first threshold as shown in Fig.~\ref{fig:polePosition} (a) and (d),  
becoming a bound and virtual state, respectively. 
The other two poles move above the two thresholds with small imaginary part,
becoming resonances, 
as shown by Fig.~\ref{fig:polePosition} (b) and (c).

\begin{figure}[h]
\centering
\includegraphics[scale=0.55]{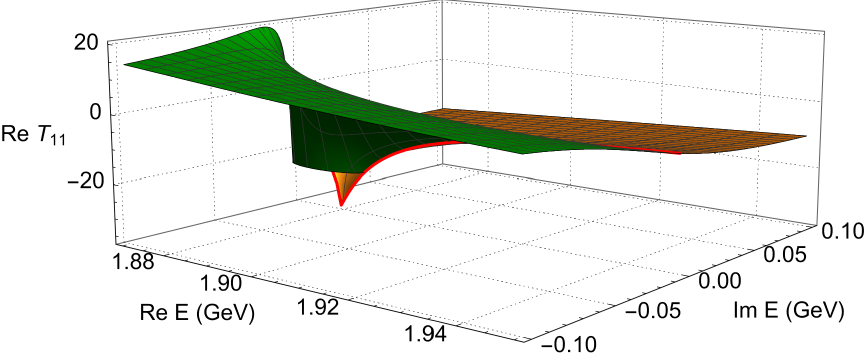} 
\includegraphics[scale=0.5]{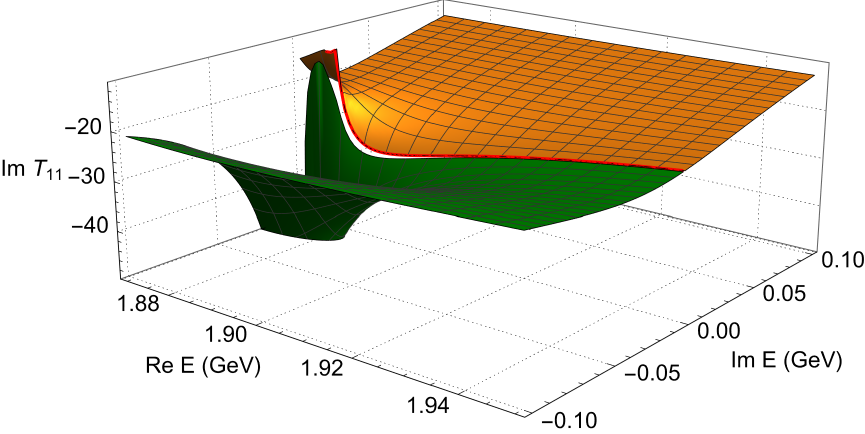} 
\caption{The real (left panel) and imaginary (right panel) parts of the scattering amplitude $T_{11}$ on the complex energy plane. The red line shows the physical range of the energy. The green and orange surfaces are the lower half plane of the third sheet and the upper half plane of the first physical sheet, respectively. 
 These two planes are connected smoothly along the real axis above the $\mathrm{thr}_2$. The $T_{21}$ scattering amplitude has the similar behavior near the $\mathrm{thr}_2$.}
\label{fig:polemm}
\end{figure}

Due to the coupled channel effects, the line shapes for the $p\bar p$ invariant mass spectrum is determined by the $T_{11}$ and $T_{21}$ behaviors. One notices that, 
the pole on the third Riemann sheet locates at 
$1868.34^{+1.66}_{-0.55}- 0.82^{+1.17}_{-1.41}i~\mathrm{MeV}$ in Solution-I. The very closeness between the $p\bar{p}$ and $n\bar{n}$ channels allows this pole significant effect on the physical lineshape (Fig.~\ref{fig:polemm}), especially the $p\bar{p}$ lineshape. Its real part is 8.20 MeV below the first threshold and 10.79 MeV below the second thresholds, respectively. Additionally, its imaginary part is sufficient small to produce a significant threshold enhancement in the $p\bar{p}$ channel as shown in Fig.~\ref{fig:lineshape}. The analogous pole for Solution-II is similar.
 The observed anomalous line shape in the $3(\pi^+\pi^-)$ channel is around $1850$ MeV, which is far below the $p \bar p$ threshold. 
 Among the four poles in both Solution-I and Solution-II, only the pole on the physical Riemann sheet is around $1850$ MeV and with a width of about $160$ MeV. This pole makes the broader peak structure in the $3(\pi^+\pi^-)$ channel. The width is even much larger than those of the $X(1840)$ and the $X(1835)$ in experiment (Tab.~\ref{tab:Xparticle}). Note that the pole positions on Riemann sheet $\{+,+\}$ $\{-,-\}$ for the two solutions are close to each other,
 which are driven by the experimental data and not beyond our expectation.

With the fitted parameters, we can also extract 
the corresponding effective couplings from the residue of the scattering amplitude
\begin{align}
\label{eq:effc}
g_i^2=\lim_{E\to E_{\mathrm{pole}}} (E-E_{\mathrm{pole}}) T_{ii}(E),
\end{align}
where $E_{\mathrm{pole}}$ is the pole position. 
$T_{ii}$ is the scattering amplitude from the $i$th channel to the $i$th channel. The effective 
couplings are collected, as shown by $g_{p\bar{p}}$ and $g_{n\bar{n}}$, in Tab.~\ref{tab:pole}. 
One can also transfer these effective 
couplings to isospin basis
\begin{align}
T^I= \mathcal R\cdot T \cdot \mathcal R^{-1},
\end{align}
by a rotation matrix
\begin{align}
    \mathcal R= \frac{1}{\sqrt 2}\left( \begin{array}{cc}
 1 & 1 \\
 1 &-1
    \end{array} \right ).
\end{align}
Here $T^I$ is the scattering amplitude in the isospin basis, which should be a diagonal matrix in the isospin limit.

With Eq.~\eqref{eq:effc}, we can obtain the effective couplings $g_{I}$ from
\begin{align}
\left( \begin{array}{cc}
 g_0^2   & g_0 g_1 \\
 g_0 g_1 & g_1^2
    \end{array} \right )  =\lim_{E\to E_{\mathrm{pole}}} (E-E_{\mathrm{pole}}) \mathcal R\cdot T\cdot\mathcal R^{-1}
 =\mathcal R\cdot\left( \begin{array}{cc}
 g_{p\bar p}^2 & g_{p\bar p} g_{n\bar n}\\
 g_{p\bar p} g_{n\bar n} &g_{n\bar n}^2
    \end{array} \right )\cdot \mathcal R^{-1}.
\end{align}
The values of $g_I$ are also listed in the Tab.~\ref{tab:pole}. One can see that only the pole on the physical sheet strongly couples to isospin singlet channel for both solutions. The other poles favor to isospin triplet channel. 
Especially, the importance of isospin triplet is also similar to that in Ref.~\cite{Kang:2015yka,Kang:2015hua}, which claim that an isospin triplet $p \bar p$ bound state $^1 S_0$ is needed to describe prominent peak shown in $J/\psi \to \gamma p \bar p$. One may naively expect that the peak structure in $\pi^0p\bar{p}$ channel is more significant, due to the $p\bar{p}$ isospin triplet property. However, there are 
various nucleon and anti-nucleon excitation in the $\pi^0 p$ and $\pi^0 \bar{p}$ channels, respectively, which makes the case more complicated~\cite{Sibirtsev:2004id}.

\begin{table}[h]
\renewcommand{\arraystretch}{1.3}
\caption{The pole positions (the second and the fifth rows) in the $\mathrm{MeV}$ unit and effective couplings in the $\mathrm{GeV}^{-1/2}$ unit (only the center values are listed) in both particle basis (the third and the sixth rows) and isospin basis (the fourth and the seventh rows) for the two solutions. The thresholds of the $p\bar p$ and $n\bar n$ are $\mathrm{thr}_1=1876.54$ MeV and $\mathrm{thr}_2=1879.13$ MeV, respectively.}
\begin{tabular}{ccccc}
\hline \hline
R. S.           & $\{+,+\}$            & $\{+,-\}$    & $\{-,+\}$        & $\{-,-\}$ \tabularnewline
\hline 
Solution-I & $1851.90^{+3.02}_{-2.71}-80.49^{+1.68}_{-1.63} i$  & $1866.07^{+22.41}_{-7.20}+86.34^{+8.65}_{-12.76}i$    
& $1857.46^{+20.61}_{-6.60} +87.20^{+9.45}_{-12.92} i$ & $1868.34^{+1.66}_{-0.55} -0.82^{+1.17}_{-1.41} i$ \tabularnewline
$(g_{p\bar p},g_{n\bar n})$ & $(1.61,1.64)$   & $(3.42,2.22)$    & $(2.16,3.42)$   & $ (0.98,0.94)$ \tabularnewline
$(g_{1},g_{0})$             & $(0.017,2.37)$  & $(3.32,2.36)$   & $(3.31,2.32)$   & $ (1.35,0.032)$ \tabularnewline
Solution-II& $1852.90^{+3.57}_{-3.31}-82.35^{+2.45}_{-2.80}i$  &$1860.31^{+8.77}_{-8.34}+61.23^{+6.18}_{-5.38}i$  
& $1855.23^{+8.49}_{-8.07} + 62.01^{+6.02}_{-5.29} i$  &$1868.92^{+1.13}_{-1.48} -2.58^{+1.70}_{-1.86} i$
\tabularnewline
$(g_{p\bar p},g_{n\bar n})$ & $(1.85,1.88)$  & $(2.82,1.80)$    & $(1.76,2.82)$ & $ (0.94,0.90)$ \tabularnewline
$(g_{1},g_{0})$             & $(0.013,2.89)$ & $(2.68,1.99)$    & $(2.67,1.98)$ & $ (1.30,0,033)$ \tabularnewline
\hline \hline
\end{tabular}
\label{tab:pole}
\end{table}

Furthermore, we can also extract scattering length and effective range from Effective-Range-Expansion (ERE) 
\begin{align}
T^{-1}(k)=-\frac{\mu}{2\pi}\left[ -\frac{1}{a_0}+\frac{1}{2}r_0 k^2-ik+\mathcal{O} \left( k^4\right) \right],
\end{align}
where $\mu=\frac{m_1 m_2}{m_1+m_2}$ is the reduce mass. $a_0$ and $r_0$ are scattering length and effective range, respectively. Thus, one can obtain 
\begin{align}
a_0&=\frac{2\pi}{\mu} T(k)|_{k\to 0},\quad
r_0=-\frac{2\pi}{\mu^2}\mathrm{Re}\left[ \frac{d T^{-1}(E)}{d E} \right ],
\end{align}
where $E=m_1+m_2+\frac{k^2}{2\mu}$. The Weinberg criterion can be used to characterise the compositeness 
\begin{align}
\bar X=\frac{1}{\sqrt{1+2|\frac{r'_0}{\mathrm{Re}[a_0]}|}}
\end{align}
of bound, virtual and resonance states~\cite{Baru:2021ldu}, 
which is always less than 1. Because that the thresholds of the $p\bar{p}$ and $n\bar{n}$ channels are very close to each other, the effective range will obtain large correction from isospin breaking effect 
\begin{align}
r'_0=r_0+\sqrt{\frac{1}{2\mu \Delta}}=r_0+\sqrt{\frac{1}{4\mu (m_n-m_p)}}
\end{align}
proposed in Ref.~\cite{Du:2021zzh} for the double charm tetraquark state.
The effective ranges, scattering lengths and compositeness are listed in Table.~\ref{tab:ar}. We can see that the compositeness $\bar X $ are about $0.98$ and $0.91$, respectively. Thus we conclude that there are $p\bar{p}$ dynamical generated state, which agrees/disagrees with those in Refs.~\cite{Datta:2003iy,Liu:2004er, Dedonder:2009bk, Kang:2015hua, Xiao:2024jmu}/Ref.~\cite{Yang:2022kpm}. 
\begin{table}[h]
\caption{The effective ranges and scattering lengths at the lowest threshold as well as the compositenesss.}
\renewcommand{\arraystretch}{1.3}
\begin{tabular}{ccccc}
\hline \hline
           & $r_0$(fm) & $r'_0$(fm) & $a_0$(fm)        & $\bar X $ \tabularnewline
\hline 
Solution-I & $-2.30$   & $1.7$    & $-65.46-31.94i$ & 0.98 \tabularnewline
Solution-II& $1.50$    & $5.50$   & $-56.52-27.16i$ & 0.91  \tabularnewline
\hline \hline
\end{tabular}
\label{tab:ar}
\end{table}

\begin{figure}[h]
\centering
\includegraphics[scale=0.8]{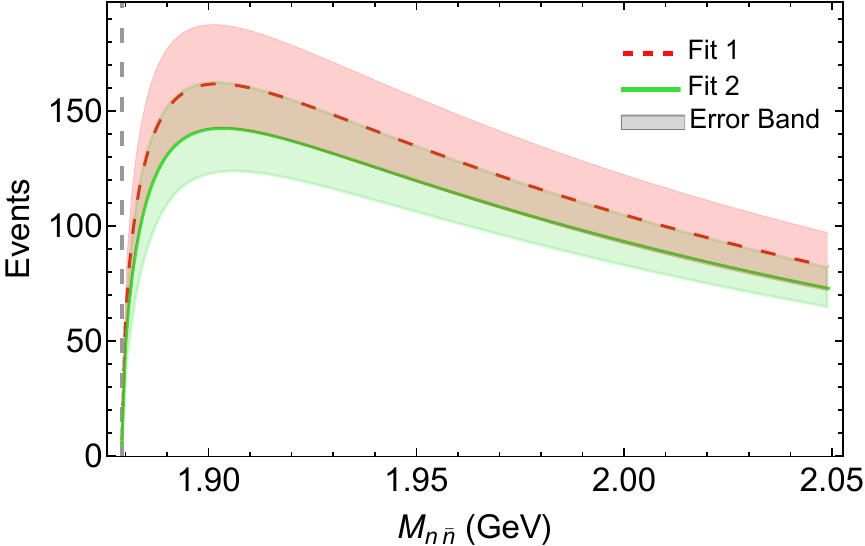} 
\caption{The predicted $n \bar n$ line shapes for the two solutions. 
The red dashed and green solid curves are for solution-I and solution-II, respectively. The error bands are one standard deviation error bands.}
\label{fig:nnlineshape}
\end{figure}

With the fitted parameters, we can also predict the $n\bar n$ line shape shown by Fig.~\ref{fig:nnlineshape}. We can also see a clear threshold enhancement in the $n\bar{n}$ channel, but not as significant as that in the $p\bar{p}$ channel, which can be used to compare with the measurement in the coming experiment. 

\section{Summary and Outlook}
We employ the chiral effective field theory and LSE to analyse the line shape of $J/\psi \to \gamma p \bar p$ and $J/\psi \to \gamma 3(\pi^+\pi^-)$. In our frame work, only the contact interactions of $N\bar N$ are involved and the leading order low energy constants are set to be as complex number to account for the $N \bar N$ annihilation. 
By performing a fit to the experimental data, we find two solutions. Their pole positions and effects on the lineshapes are analogous. 
It shows that the structures near the $p \bar p$ threshold are because of the pole on the Riemann sheet $\{-,-\}$ with mass about $1868~\mathrm{MeV}$ and width $2~\mathrm{MeV}$. This pole more favors to couple to isospin triplet channel than that to isospin singlet channel.
It also assign the broader bump structure in the $J/\psi \to \gamma 3(\pi^+\pi^-)$ process to the pole on the physical Riemann sheet with mass about $1850$ MeV and width about $160$ MeV. This pole more favors to couple to isospin singlet channel channel than that to isospin triplet channel,
in contrary to the pole mentioned above. This kind of property gives us an intuition that in which channel and where we would expect a peak structure. That the pole positions of the two solutions are almost the same is not beyond our expectation, as they should be driven by the experimental lineshapes. 
With the fitted parameters, 
we also obtain the compositeness of the two solutions,
i.e. $0.98$ and $0.91$ for Solution-I and Solution-II, respectively. Those large values suggest that there 
should exit $p\bar{p}$ resonance based on the current experimental data. 
In addition, we also see a clear threshold enhancement in the $n\bar{n}$ channel, but not as significant as that in $p\bar{p}$ channel, which is useful and compared with further experimental measurement.

\begin{acknowledgments}
We are grateful to Tong Chen, Shuangshi Fang and Jifeng Hu for the very helpful discussion. This work is partly supported by the National Natural Science Foundation of China with Grant Nos.~12375073, ~12035007, ~12147128, ~12235018,  Guangdong Provincial funding with Grant No.~2019QN01X172, Guangdong Major Project of Basic and Applied Basic Research No.~2020B0301030008. Q.W. and Q.Z. are also supported by the NSFC and the Deutsche Forschungsgemeinschaft (DFG, German Research Foundation) through the funds provided to the Sino-German Collaborative Research Center TRR110 ``Symmetries and the Emergence of Structure in QCD" (NSFC Grant No. 12070131001, DFG Project-ID 196253076-TRR 110). Q.Z. is also supported by National Key Basic Research Program of China under
Contract No. 2020YFA0406300, and Strategic Priority
Research Program of Chinese Academy of Sciences
(Grant No. XDB34030302).
\end{acknowledgments}

\begin{appendix}

\section{The partial wave projection}
The partial wave amplitudes can be obtained with
\begin{align}
\label{eq:PWA}
&T^{J,J_z}_{L,S;L',S'}=\langle J,J_z;L',S'| \hat T |J,J_z;L,S\rangle \notag \\
&= \frac{1}{2J+1} \frac{1}{\sqrt{N_i!}}\frac{1}{\sqrt{N_f!}}\sum_{s_1^z,s_2^z,L_z}  \sum_{s_3^z,s_4^z,L_z'}
\langle s_1,s_1^z;s_2,s_2^z | S,S_z \rangle  \langle L,L_z;S,S_z |J,J_z\rangle
\langle s_3,s_3^z;s_4,s_4^z | S',S_z' \rangle  \langle L',L_z';S',S'_z |J,J_z\rangle \notag \\
&\times \int d\Omega_{\bm p'} Y_{L,L_z}(0,0) Y_{L',L'_z}^*(\Omega_{\bm p'})
\langle \bm p'(s_3,s^z_3); -\bm p'(s_4,s^z_4) | \hat T|\bm p(s_1,s^z_1);-\bm p(s_2,s^z_2)\rangle,
\end{align}
where $J (J_z)$ is the total angular momentum (the third component) of the two-body system. $N_i(N_f)$ is the number of identical particles of initial(final) states.
$S(S')$ and $L(L')$ are the total spin and the orbital angular momentum of the incoming and outgoing two-body systems, respectively. 
$|\bm p(s_1,s^z_1);-\bm p(s_2,s^z_2)\rangle$ is the canonical state of the two hadron system. $s_i$ and $s^{z}_i$ are the spin and the third component of the $i$-th particle, respectively.

\section{Kinematics}
\label{app:kinematics}
\subsection{3-body decay}

\begin{figure*}[h]
\centering
\includegraphics[scale=0.5]{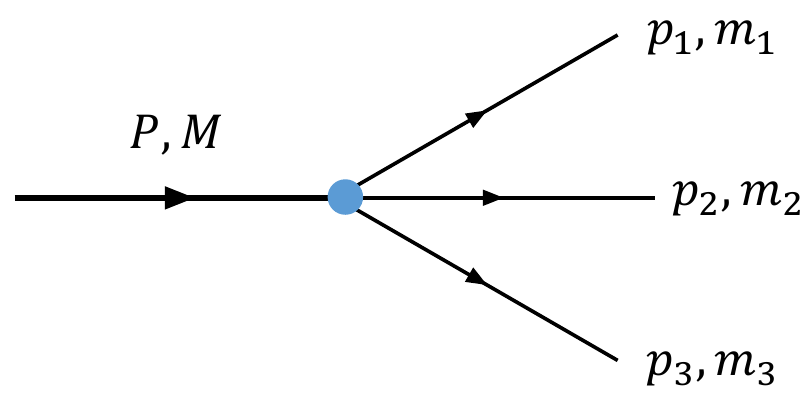} 
\caption{Definitions of variables for 3-body decays.}
\label{fig:3body}
\end{figure*}

Defining $p_{ij}=p_i+p_j$ and $s_{ij}=m^2_{ij}=p^2_{ij}$, the energy and momentum of the final stats can be expressed as 
\begin{align}
&E_1=\frac{M^2+m_1^2-m_{23}^2}{2M},
&&E_2=\frac{M^2+m_2^2-m_{13}^2}{2M}, 
&&E_3=\frac{M^2+m_3^2-m_{12}^2}{2M}, \notag\\  
&|\bm p_1|=\frac{\lambda^{1/2}(M^2,m_1^2,m_{23}^2)}{2M}, 
&&|\bm p_2|=\frac{\lambda^{1/2}(M^2,m_2^2,m_{13}^2)}{2M},
&&|\bm p_3|=\frac{\lambda^{1/2}(M^2,m_3^2,m_{12}^2)}{2M}, \notag
\end{align}
where the K\"all\'en function $\lambda$ is defined as
\begin{align}
\lambda(a,b,c)&=a^2+b^2+c^2-2ab-2ac-2bc \notag \\
&=\left[a-(\sqrt{b}-\sqrt{c})^2\right]\left[a-(\sqrt{b}+\sqrt{c})^2\right].
\end{align}
Note that $m^2_{ij}$ satisfy that
\begin{align}
m^2_{12}+m^2_{23}+m^2_{13}=M^2+m_1^2+m_2^2+m_3^2.
\end{align}
This means that only two of them are independent and the range of $m_{ij}^2$ is
\begin{align}
m_{ij}^2\in \left[(m_i+m_j)^2,(M-m_k)^2\right ],~(i\neq j \neq k).
\end{align}
The differential decay rate for the 3-body decay processes can be written in the form~\cite{Workman:2022ynf}
\begin{align}
d \Gamma= \frac{1}{(2\pi)^3}\frac{1}{32 M^3}\frac{1}{2S +1}|\mathcal M|^2 d m^2_{12} d m^2_{23}.
\end{align}
The upper and lower limits of $m^2_{12}$ and $m^2_{23}$ are
\begin{align}
(m^2_{12})_\mathrm{min}&=(m_1+m_2)^2,\quad (m^2_{12})_\mathrm{max}=(M-m_3)^2, \notag \\
\left(m_{23}^2\right)_{\min }&= 
\left(E_2^*+E_3^*\right)^2-\left(\sqrt{E_2^{* 2}-m_2^2}+\sqrt{E_3^{* 2}-m_3^2}\right)^2 \notag \\
&= \left(E_2^*+E_3^*\right)^2-\frac{1}{4m^2_{12}} \left[\lambda^{1/2}(m^2_{12},m_1^2,m_2^2) + \lambda^{1/2}(M^2,m_{12}^2,m_3^2) \right]^2, \\
\left(m_{23}^2\right)_{\max }&=
\left(E_2^*+E_3^*\right)^2-\left(\sqrt{E_2^{* 2}-m_2^2}-\sqrt{E_3^{* 2}-m_3^2}\right)^2 \notag\\
&= \left(E_2^*+E_3^*\right)^2-\frac{1}{4m^2_{12}} \left[\lambda^{1/2}(m^2_{12},m_1^2,m_2^2) - \lambda^{1/2}(M^2,m_{12}^2,m_3^2) \right]^2.
\end{align}
Here 
\begin{align}
E_2^*=\frac{m_{12}^2-m_1^2+m_2^2}{2m_{12}},\quad E_3^*=\frac{M^2-m_{12}^2-m_3^2}{2m_{12}}
\end{align}
are the energies of particles 2 and 3 in the $m_{12}$ rest frame.

\subsection{4-body decay}
\begin{figure*}[h]
\centering
\includegraphics[scale=0.5]{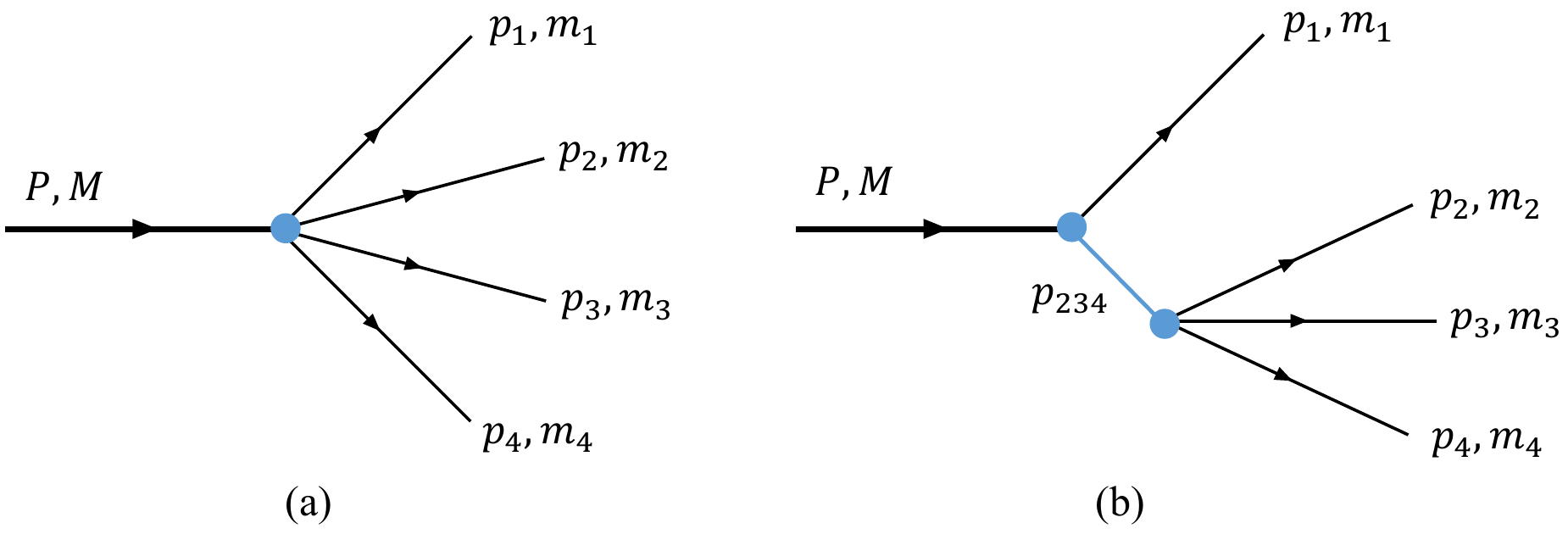} 
\caption{Definitions of variables for 4-body decays.}
\label{fig:4body}
\end{figure*}

There are many work~\cite{Byckling:1969sx,Asatrian:2012tp,Jing:2020tth,Yu:2021dtp} devote to perform the calculation of four body or $n$-body phase space. Here we list the formulas used in this work.
For the 4-body decay process, the phase space is given by
\begin{align}
\label{eq:space}
\int d\Phi_4&=\int  (2\pi)^4\delta^4(P-\sum_{i=1}^4 p_i)\prod_{i=1}^4 \frac{d^3\bm p_i}{(2\pi)^3 2E_i}.
\end{align}
We insert two integrals with $p_{234}=p_2+p_3+p_4$,
\begin{align}
1&=\int \frac{d^4 p_{234}}{(2\pi)^4} (2\pi)^4\delta^4(p_{234}-p_2-p_3-p_4) \theta_0(p_{234}^0), \notag \\
1&=\int \frac{dm^2_{234}}{2\pi} 2\pi\delta(m^2_{234}-p^2_{234}).
\end{align}
With these two integrals, we can obtain that
\begin{align}
1&=\int \frac{dm^2_{234}}{2\pi} \frac{d^4 p_{234}}{(2\pi)^4} (2\pi)^4\delta^4(p_{234}-p_2-p_3-p_4) \theta_0(p_{234}^0)  2\pi\delta(m^2_{234}-p^2_{234})\notag  \\
&=\int \frac{dm^2_{234}}{2\pi} \frac{d^3\bm p_{234}}{(2\pi)^3 2E_{234}} (2\pi)^4\delta^4(p_{234}-p_2-p_3-p_4).
\end{align}
$E_{234}$ satisfies $p_{234}^2=E^2_{234}-\bm p_{234}^2$.
With these results, the 4-body phase space can be written as
\begin{align}
d\Phi_4&=\int d\Phi_4 \frac{dm^2_{234}}{2\pi} \frac{d^3\bm p_{234}}{(2\pi)^3 2E_{234}} (2\pi)^4\delta^4(p_{234}-p_2-p_3-p_4) =\int \frac{d m^2_{234}}{2\pi} d\Phi_2 d\Phi_3,
\end{align}
where 
\begin{align}
d\Phi_2&=(2\pi)^4 \delta^4(P-p_1-p_{234})\frac{d^3\bm p_1}{(2\pi)^3 2E_1} \frac{d^3\bm p_{234}}{(2\pi)^3 2 E_{234}}, \\
d\Phi_3&=(2\pi)^4 \delta^4(p_{234}-p_1-p_2-p_3)\prod_{i=2}^4 \frac{d^3\bm p_i}{(2\pi)^3 2E_i}.
\end{align}
This result means that the 4-body phase space can be decomposed into 2-body part and 3-body part as illustrated by Fig.~\ref{fig:4body}(b).
$d\Phi_2$ and $d\Phi_3$ are all Lorentz invariant, thus we can perform the calculation in any frame.

With the 2-body and 3-body phase space formulas, the 4-body phase space can be reduced as
\begin{align}
\int d\Phi_4 =\frac{4\pi}{(2\pi)^2} \frac{|\bm p_1|}{4 M} 
\times \frac{1}{16 (2\pi)^3 m^2_{234}}\int_{s_{23}^-}^{s_{23}^+} d s_{23} \int_{s_{34}^-}^{s_{34}^+} d s_{34} \int \frac{d m^2_{234}}{2\pi},
\end{align}
where
\begin{align}
|\bm p_1|=\frac{\sqrt{[M^2-(m_1+m_{234})^2][M^2-(m_1-m_{234})^2]}}{2 M}
\end{align}
is the momentum of particle 1 in the rest frame of initial rest frame. Here we have taken the assumptions that the amplitude is independent of all the angle but depends on $m^2_{234}$. The upper and lower limits $s_{23,34}^{\pm}$ are
\begin{align}
s_{23}^-&=(m_2+m_3)^2,\quad s_{23}^+=(m_{234}-m_4)^2, \\
s_{34}^\pm&=\frac{1}{4m_{23}}\left[ (m^2_{234}-m_2^2-m_4^2+m_3^2)^2 - \left(\lambda^{1/2}(m_{23}^2,m_2^2,m_3^2)\mp \lambda^{1/2}(m^2_{234},m_{23}^2,m_4^2)\right)^2\right].
\end{align}

In the $3(\pi^+\pi^-)$ rest frame, the energy of the particles are expressed as
\begin{align}
E^*_{M}&=\frac{m^2_{M}+s_{234}}{2 \sqrt{s_{234}}}, \\
E_2^* & =\frac{s_{234}-s_{34}+m_2^2}{2 \sqrt{s_{234}}}, \\
E_3^* & =\frac{s_{234}-s_{24}+m_3^2}{2 \sqrt{s_{234}}}, \\
E_4^* & =\frac{s_{234}-s_{23}+m_4^2}{2 \sqrt{s_{234}}}, \\
E^*_1 & =E^*_{M}-\sqrt{s_{234}}=\frac{m_{M}^2-s_{234}}{2 \sqrt{s_{234}}},
\end{align}
where $s_{234}=m^2_{234}=p^2_{234}=(p_2+p_3+p_4)^2$.

\end{appendix}


\begin{thebibliography}{100}
\bibitem{Fermi:1949voc}
E.~Fermi and C.~N.~Yang,
Phys. Rev. \textbf{76}, 1739-1743 (1949)
doi:10.1103/PhysRev.76.1739

\bibitem{Nambu:1961tp}
Y.~Nambu and G.~Jona-Lasinio,
Phys. Rev. \textbf{122}, 345-358 (1961)
doi:10.1103/PhysRev.122.345

\bibitem{Nambu:1961fr}
Y.~Nambu and G.~Jona-Lasinio,
Phys. Rev. \textbf{124}, 246-254 (1961)
doi:10.1103/PhysRev.124.246

\bibitem{BES:2003aic}
J.~Z.~Bai \textit{et al.} [BES],
Phys. Rev. Lett. \textbf{91}, 022001 (2003)
doi:10.1103/PhysRevLett.91.022001
[arXiv:hep-ex/0303006 [hep-ex]].

\bibitem{BES:2005ega}
M.~Ablikim \textit{et al.} [BES],
Phys. Rev. Lett. \textbf{95}, 262001 (2005)
doi:10.1103/PhysRevLett.95.262001
[arXiv:hep-ex/0508025 [hep-ex]].

\bibitem{BES:2007inv}
M.~Ablikim \textit{et al.} [BES],
Eur. Phys. J. C \textbf{53}, 15-20 (2008)
doi:10.1140/epjc/s10052-007-0467-4
[arXiv:0710.5369 [hep-ex]].

\bibitem{BESIII:2010vwa}
M.~Ablikim \textit{et al.} [BESIII],
Chin. Phys. C \textbf{34}, 421
doi:10.1088/1674-1137/34/4/001
[arXiv:1001.5328 [hep-ex]].

\bibitem{CLEO:2010fre}
J.~P.~Alexander \textit{et al.} [CLEO],
Phys. Rev. D \textbf{82}, 092002 (2010)
doi:10.1103/PhysRevD.82.092002
[arXiv:1007.2886 [hep-ex]].

\bibitem{BES:2009ufh}
M.~Ablikim \textit{et al.} [BES],
Phys. Rev. D \textbf{80}, 052004 (2009)
doi:10.1103/PhysRevD.80.052004
[arXiv:0905.1562 [hep-ex]].

\bibitem{BESIII:2010gmv}
M.~Ablikim \textit{et al.} [BESIII],
Phys. Rev. Lett. \textbf{106}, 072002 (2011)
doi:10.1103/PhysRevLett.106.072002
[arXiv:1012.3510 [hep-ex]].

\bibitem{BESIII:2011aa}
M.~Ablikim \textit{et al.} [BESIII],
Phys. Rev. Lett. \textbf{108}, 112003 (2012)
doi:10.1103/PhysRevLett.108.112003
[arXiv:1112.0942 [hep-ex]].

\bibitem{BESIII:2013sbm}
M.~Ablikim \textit{et al.} [BESIII],
Phys. Rev. D \textbf{88}, no.9, 091502 (2013)
doi:10.1103/PhysRevD.88.091502
[arXiv:1305.5333 [hep-ex]].

\bibitem{BESIII:2013lac}
M.~Ablikim \textit{et al.} [BESIII],
Phys. Rev. D \textbf{87}, no.11, 112004 (2013)
doi:10.1103/PhysRevD.87.112004
[arXiv:1303.3108 [hep-ex]].

\bibitem{BESIII:2016fbr}
M.~Ablikim \textit{et al.} [BESIII],
Phys. Rev. Lett. \textbf{117}, no.4, 042002 (2016)
doi:10.1103/PhysRevLett.117.042002
[arXiv:1603.09653 [hep-ex]].

\bibitem{BESIII:2023vvr}
M.~Ablikim \textit{et al.} [BESIII],
Phys. Rev. Lett. \textbf{132}, no.15, 151901 (2024)
doi:10.1103/PhysRevLett.132.151901
[arXiv:2310.17937 [hep-ex]].

\bibitem{BaBar:2005pon}
B.~Aubert \textit{et al.} [BaBar],
Phys. Rev. D \textbf{73}, 012005 (2006)
doi:10.1103/PhysRevD.73.012005
[arXiv:hep-ex/0512023 [hep-ex]].

\bibitem{BaBar:2013ves}
J.~P.~Lees \textit{et al.} [BaBar],
Phys. Rev. D \textbf{87}, no.9, 092005 (2013)
doi:10.1103/PhysRevD.87.092005
[arXiv:1302.0055 [hep-ex]].

\bibitem{CMD-3:2015fvi}
R.~R.~Akhmetshin \textit{et al.} [CMD-3],
Phys. Lett. B \textbf{759}, 634-640 (2016)
doi:10.1016/j.physletb.2016.04.048
[arXiv:1507.08013 [hep-ex]].

\bibitem{Belle:2002bro}
K.~Abe \textit{et al.} [Belle],
Phys. Rev. Lett. \textbf{88}, 181803 (2002)
doi:10.1103/PhysRevLett.88.181803
[arXiv:hep-ex/0202017 [hep-ex]].

\bibitem{Belle:2002fay}
K.~Abe \textit{et al.} [Belle],
Phys. Rev. Lett. \textbf{89}, 151802 (2002)
doi:10.1103/PhysRevLett.89.151802
[arXiv:hep-ex/0205083 [hep-ex]].

\bibitem{Belle:2007oni}
J.~T.~Wei \textit{et al.} [Belle],
Phys. Lett. B \textbf{659}, 80-86 (2008)
doi:10.1016/j.physletb.2007.11.063
[arXiv:0706.4167 [hep-ex]].

\bibitem{BaBar:2007fsu}
B.~Aubert \textit{et al.} [BaBar],
Phys. Rev. D \textbf{76}, 092006 (2007)
doi:10.1103/PhysRevD.76.092006
[arXiv:0709.1988 [hep-ex]].

\bibitem{CMD-3:2013nph}
R.~R.~Akhmetshin \textit{et al.} [CMD-3],
Phys. Lett. B \textbf{723}, 82-89 (2013)
doi:10.1016/j.physletb.2013.04.065
[arXiv:1302.0053 [hep-ex]].

\bibitem{Milstein:2022tfx}
A.~I.~Milstein and S.~G.~Salnikov,
Phys. Rev. D \textbf{106}, no.7, 074012 (2022)
doi:10.1103/PhysRevD.106.074012
[arXiv:2207.14020 [hep-ph]].

\bibitem{Gao:2003ka}
C.~S.~Gao and S.~L.~Zhu,
Commun. Theor. Phys. \textbf{42}, 844 (2004)
doi:10.1088/0253-6102/42/6/844
[arXiv:hep-ph/0308205 [hep-ph]].

\bibitem{Datta:2003iy}
A.~Datta and P.~J.~O'Donnell,
Phys. Lett. B \textbf{567}, 273-276 (2003)
doi:10.1016/j.physletb.2003.06.050
[arXiv:hep-ph/0306097 [hep-ph]].

\bibitem{Yan:2004xs}
M.~L.~Yan, S.~Li, B.~Wu and B.~Q.~Ma,
Phys. Rev. D \textbf{72}, 034027 (2005)
doi:10.1103/PhysRevD.72.034027
[arXiv:hep-ph/0405087 [hep-ph]].

\bibitem{Liu:2004er}
X.~Liu, X.~Q.~Zeng, Y.~B.~Ding, X.~Q.~Li, H.~Shen and P.~N.~Shen,
[arXiv:hep-ph/0406118 [hep-ph]].

\bibitem{Ding:2007xi}
G.~J.~Ding and M.~L.~Yan,
Phys. Rev. C \textbf{75}, 034004 (2007)
doi:10.1103/PhysRevC.75.034004
[arXiv:nucl-th/0702037 [nucl-th]].

\bibitem{Deng:2012wi}
C.~Deng, J.~Ping, Y.~Yang and F.~Wang,
Phys. Rev. D \textbf{86}, 014008 (2012)
doi:10.1103/PhysRevD.86.014008
[arXiv:1202.4167 [hep-ph]].

\bibitem{Dedonder:2009bk}
J.~P.~Dedonder, B.~Loiseau, B.~El-Bennich and S.~Wycech,
Phys. Rev. C \textbf{80}, 045207 (2009)
doi:10.1103/PhysRevC.80.045207
[arXiv:0904.2163 [nucl-th]].

\bibitem{Liu:2009vm}
X.~H.~Liu, Y.~J.~Zhang and Q.~Zhao,
Phys. Rev. D \textbf{80}, 034032 (2009)
doi:10.1103/PhysRevD.80.034032
[arXiv:0903.1427 [hep-ph]].

\bibitem{Ding:2005gh}
G.~J.~Ding and M.~L.~Yan,
Eur. Phys. J. A \textbf{28}, 351-360 (2006)
doi:10.1140/epja/i2006-10055-3
[arXiv:hep-ph/0511186 [hep-ph]].

\bibitem{Wang:2006sna}
Z.~G.~Wang and S.~L.~Wan,
J. Phys. G \textbf{34}, 505-511 (2007)
doi:10.1088/0954-3899/34/3/008
[arXiv:hep-ph/0601105 [hep-ph]].

\bibitem{Liu:2007tj}
C.~Liu,
Eur. Phys. J. C \textbf{53}, 413-419 (2008)
doi:10.1140/epjc/s10052-007-0471-8
[arXiv:0710.4185 [hep-ph]].

\bibitem{Wycech:2015qra}
S.~Wycech, J.~P.~Dedonder and B.~Loiseau,
Hyperfine Interact. \textbf{234}, no.1-3, 141-148 (2015)
doi:10.1007/s10751-015-1150-z
[arXiv:1503.03319 [nucl-th]].

\bibitem{Chen:2011yu}
G.~Y.~Chen and J.~P.~Ma,
Phys. Rev. D \textbf{83}, 094029 (2011)
doi:10.1103/PhysRevD.83.094029
[arXiv:1101.4071 [hep-ph]].

\bibitem{Chen:2010an}
G.~Y.~Chen, H.~R.~Dong and J.~P.~Ma,
Phys. Lett. B \textbf{692}, 136-142 (2010)
doi:10.1016/j.physletb.2010.07.030
[arXiv:1004.5174 [hep-ph]].

\bibitem{Chen:2008ee}
G.~Y.~Chen, H.~R.~Dong and J.~P.~Ma,
Phys. Rev. D \textbf{78}, 054022 (2008)
doi:10.1103/PhysRevD.78.054022
[arXiv:0806.4661 [hep-ph]].

\bibitem{Kang:2015yka}
X.~W.~Kang, J.~Haidenbauer and U.~G.~Mei\ss{}ner,
Phys. Rev. D \textbf{91}, no.7, 074003 (2015)
doi:10.1103/PhysRevD.91.074003
[arXiv:1502.00880 [nucl-th]].

\bibitem{Milstein:2017dqp}
A.~I.~Milstein and S.~G.~Salnikov,
Nucl. Phys. A \textbf{966}, 54-63 (2017)
doi:10.1016/j.nuclphysa.2017.06.002
[arXiv:1701.05741 [hep-ph]].

\bibitem{Salnikov:2023ipo}
S.~G.~Salnikov and A.~I.~Milstein,
Nucl. Phys. B \textbf{1002}, 116539 (2024)
doi:10.1016/j.nuclphysb.2024.116539
[arXiv:2311.14309 [hep-ph]].

\bibitem{Li:2005vd}
B.~A.~Li,
Phys. Rev. D \textbf{74}, 034019 (2006)
doi:10.1103/PhysRevD.74.034019
[arXiv:hep-ph/0510093 [hep-ph]].

\bibitem{Kochelev:2005tu}
N.~Kochelev and D.~P.~Min,
Phys. Rev. D \textbf{72}, 097502 (2005)
doi:10.1103/PhysRevD.72.097502
[arXiv:hep-ph/0510016 [hep-ph]].

\bibitem{Kochelev:2005vd}
N.~Kochelev and D.~P.~Min,
Phys. Lett. B \textbf{633}, 283-288 (2006)
doi:10.1016/j.physletb.2005.11.079
[arXiv:hep-ph/0508288 [hep-ph]].

\bibitem{He:2005nm}
X.~G.~He, X.~Q.~Li, X.~Liu and J.~P.~Ma,
Eur. Phys. J. C \textbf{49}, 731-736 (2007)
doi:10.1140/epjc/s10052-006-0129-y
[arXiv:hep-ph/0509140 [hep-ph]].

\bibitem{Hao:2005hu}
G.~Hao, C.~F.~Qiao and A.~L.~Zhang,
Phys. Lett. B \textbf{642}, 53-61 (2006)
doi:10.1016/j.physletb.2006.09.031
[arXiv:hep-ph/0512214 [hep-ph]].

\bibitem{Gui:2019dtm}
L.~C.~Gui, J.~M.~Dong, Y.~Chen and Y.~B.~Yang,
Phys. Rev. D \textbf{100}, no.5, 054511 (2019)
doi:10.1103/PhysRevD.100.054511
[arXiv:1906.03666 [hep-lat]].

\bibitem{Huang:2005bc}
T.~Huang and S.~L.~Zhu,
Phys. Rev. D \textbf{73}, 014023 (2006)
doi:10.1103/PhysRevD.73.014023
[arXiv:hep-ph/0511153 [hep-ph]].

\bibitem{Yu:2011ta}
J.~S.~Yu, Z.~F.~Sun, X.~Liu and Q.~Zhao,
Phys. Rev. D \textbf{83}, 114007 (2011)
doi:10.1103/PhysRevD.83.114007
[arXiv:1104.3064 [hep-ph]].

\bibitem{Wang:2020due}
L.~M.~Wang, Q.~S.~Zhou, C.~Q.~Pang and X.~Liu,
Phys. Rev. D \textbf{102}, no.11, 114034 (2020)
doi:10.1103/PhysRevD.102.114034
[arXiv:2010.05132 [hep-ph]].

\bibitem{Li:2008mza}
D.~M.~Li and B.~Ma,
Phys. Rev. D \textbf{77}, 074004 (2008)
doi:10.1103/PhysRevD.77.074004
[arXiv:0801.4821 [hep-ph]].

\bibitem{Liu:2016yfr}
Y.~F.~Liu and X.~W.~Kang,
Symmetry \textbf{8}, no.3, 14 (2016)
doi:10.3390/sym8030014

\bibitem{Ma:2024gsw}
B.~Q.~Ma,
doi:10.1360/TB-2024-0578
[arXiv:2406.19180 [hep-ph]].

\bibitem{Workman:2022ynf}
R.~L.~Workman \textit{et al.} [Particle Data Group],
PTEP \textbf{2022}, 083C01 (2022)
doi:10.1093/ptep/ptac097

\bibitem{BESIII:2015xco}
M.~Ablikim \textit{et al.} [BESIII],
Phys. Rev. Lett. \textbf{115}, no.9, 091803 (2015)
doi:10.1103/PhysRevLett.115.091803
[arXiv:1506.04807 [hep-ex]].

\bibitem{Sibirtsev:2004id}
A.~Sibirtsev, J.~Haidenbauer, S.~Krewald, U.~G.~Meissner and A.~W.~Thomas,
Phys. Rev. D \textbf{71}, 054010 (2005)
doi:10.1103/PhysRevD.71.054010
[arXiv:hep-ph/0411386 [hep-ph]].

\bibitem{Kang:2013uia}
X.~W.~Kang, J.~Haidenbauer and U.~G.~Mei\ss{}ner,
JHEP \textbf{02}, 113 (2014)
doi:10.1007/JHEP02(2014)113
[arXiv:1311.1658 [hep-ph]].

\bibitem{Dai:2017ont}
L.~Y.~Dai, J.~Haidenbauer and U.~G.~Mei\ss{}ner,
JHEP \textbf{07}, 078 (2017)
doi:10.1007/JHEP07(2017)078
[arXiv:1702.02065 [nucl-th]].

\bibitem{Xiao:2024jmu}
Y.~Xiao, J.~X.~Lu and L.~S.~Geng,
[arXiv:2406.01292 [nucl-th]].

\bibitem{Cote:1982gr}
J.~Cote, M.~Lacombe, B.~Loiseau, B.~Moussallam and R.~Vinh Mau,
Phys. Rev. Lett. \textbf{48}, 1319 (1982)
doi:10.1103/PhysRevLett.48.1319

\bibitem{Pignone:1994uj}
M.~Pignone, M.~Lacombe, B.~Loiseau and R.~Vinh Mau,
Phys. Rev. C \textbf{50}, 2710-2730 (1994)
doi:10.1103/PhysRevC.50.2710

\bibitem{El-Bennich:1999zbs}
B.~El-Bennich, M.~Lacombe, B.~Loiseau and R.~Vinh Mau,
Phys. Rev. C \textbf{59}, 2313-2315 (1999)
doi:10.1103/PhysRevC.59.2313

\bibitem{Entem:2007bb}
D.~R.~Entem and F.~Fernandez,
Phys. Rev. D \textbf{75}, 014004 (2007)
doi:10.1103/PhysRevD.75.014004

\bibitem{Richard:2022tpn}
J.~M.~Richard,
doi:10.1007/978-981-15-8818-1\_53-2
[arXiv:2205.02529 [nucl-th]].

\bibitem{Epelbaum:2004fk}
E.~Epelbaum, W.~Glockle and U.~G.~Meissner,
Nucl. Phys. A \textbf{747}, 362-424 (2005)
doi:10.1016/j.nuclphysa.2004.09.107
[arXiv:nucl-th/0405048 [nucl-th]].

\bibitem{Yang:2022kpm}
Q.~H.~Yang, D.~Guo and L.~Y.~Dai,
Phys. Rev. D \textbf{107}, no.3, 034030 (2023)
doi:10.1103/PhysRevD.107.034030
[arXiv:2209.10101 [hep-ph]].

\bibitem{Dai:2018tlc}
L.~Y.~Dai, J.~Haidenbauer and U.~G.~Mei\ss{}ner,
Phys. Rev. D \textbf{98}, no.1, 014005 (2018)
doi:10.1103/PhysRevD.98.014005
[arXiv:1804.07077 [hep-ph]].

\bibitem{Kang:2015hua}
X.~W.~Kang,
PoS \textbf{CD15}, 108 (2016)
doi:10.22323/1.253.0108
[arXiv:1508.06189 [hep-ph]].

\bibitem{BESIII:2017qwj}
M.~Ablikim \textit{et al.} [BESIII],
Phys. Lett. B \textbf{771}, 45-51 (2017)
doi:10.1016/j.physletb.2017.05.033
[arXiv:1701.04198 [hep-ex]].

\bibitem{James:1975dr}
F.~James and M.~Roos,
Comput. Phys. Commun. \textbf{10}, 343-367 (1975)
doi:10.1016/0010-4655(75)90039-9

\bibitem{Kato:1965iee}
M.~Kato,
Annals Phys. \textbf{31}, no.1, 130-147 (1965)
doi:10.1016/0003-4916(65)90235-6

\bibitem{Badalian:1981xj}
A.~M.~Badalian, L.~P.~Kok, M.~I.~Polikarpov and Y.~A.~Simonov,
Phys. Rept. \textbf{82}, 31-177 (1982)
doi:10.1016/0370-1573(82)90014-X

\bibitem{Baru:2019xnh}
V.~Baru, E.~Epelbaum, A.~A.~Filin, C.~Hanhart, A.~V.~Nefediev and Q.~Wang,
Phys. Rev. D \textbf{99}, no.9, 094013 (2019)
doi:10.1103/PhysRevD.99.094013
[arXiv:1901.10319 [hep-ph]].

\bibitem{ParticleDataGroup:2022pth}
R.~L.~Workman \textit{et al.} [Particle Data Group],
PTEP \textbf{2022}, 083C01 (2022)
doi:10.1093/ptep/ptac097

\bibitem{Baru:2021ldu}
V.~Baru, X.~K.~Dong, M.~L.~Du, A.~Filin, F.~K.~Guo, C.~Hanhart, A.~Nefediev, J.~Nieves and Q.~Wang,
Phys. Lett. B \textbf{833}, 137290 (2022)
doi:10.1016/j.physletb.2022.137290
[arXiv:2110.07484 [hep-ph]].

\bibitem{Du:2021zzh}
M.~L.~Du, V.~Baru, X.~K.~Dong, A.~Filin, F.~K.~Guo, C.~Hanhart, A.~Nefediev, J.~Nieves and Q.~Wang,
Phys. Rev. D \textbf{105}, no.1, 014024 (2022)
doi:10.1103/PhysRevD.105.014024
[arXiv:2110.13765 [hep-ph]].


\bibitem{Byckling:1969sx}
E.~Byckling and K.~Kajantie,
Nucl. Phys. B \textbf{9}, 568-576 (1969)
doi:10.1016/0550-3213(69)90271-5


\bibitem{Asatrian:2012tp}
H.~M.~Asatrian, A.~Hovhannisyan and A.~Yeghiazaryan,
Phys. Rev. D \textbf{86}, 114023 (2012)
doi:10.1103/PhysRevD.86.114023
[arXiv:1210.7939 [hep-ph]].


\bibitem{Jing:2020tth}
H.~J.~Jing, C.~W.~Shen and F.~K.~Guo,
Science Bulletin \textbf{66}, no.7, 653-656 (2021)
doi:10.1016/j.scib.2020.10.009
[arXiv:2005.01942 [hep-ph]].


\bibitem{Yu:2021dtp}
K.~Yu, D.~M.~Li and J.~J.~Wu,
Chin. Phys. C \textbf{46}, no.8, 083101 (2022)
doi:10.1088/1674-1137/ac6666
[arXiv:2111.08901 [hep-ph]].

\end{thebibliography}

\end{document}